\documentclass[12pt, a4paper, twocolumn]{article}
\usepackage{xurl}
\usepackage[english]{babel}
\usepackage[super,comma,sort&compress]{natbib}
\usepackage[utf8]{inputenc}
\usepackage{amsthm}
\usepackage[pdftex]{hyperref}
\usepackage{amsmath}
\usepackage{comment}
\usepackage{mathtools}
\usepackage{enumitem}   
\usepackage{dsfont}
\usepackage{graphicx}
\usepackage{cancel}
\usepackage{mathrsfs}
\usepackage{xcolor}
\usepackage{abstract}
\usepackage{tensor}
\usepackage{mhchem}
\usepackage{booktabs}

\usepackage{amssymb}
\usepackage{siunitx}
\DeclareMathOperator{\arcsinh}{arcsinh}

\usepackage{hyperref}
\hypersetup{colorlinks=true, urlcolor=blue, linkcolor=blue, citecolor=blue}

\begin{document}

\title{White Dwarf mass-radius relation}

\author{Federico Ambrosino\\
	Department of Physics\\
	University of Pisa\\
	Pisa, Italia \\
    \href{mailto:f.ambrosino@studenti.unipi.it}{f.ambrosino@studenti.unipi.it}
}

\newcommand{\abstractText}{\noindent
This paper aims to provide an introduction to the problem of determining the mass-radius relationship for white dwarfs. Both Newtonian and general relativistic case are examined, electromagnetic interactions are considered and 
the problem of $\beta$-decay equilibrium is discussed in particular for Helium WD.}

\twocolumn[
  \begin{@twocolumnfalse}
    \maketitle
    \begin{abstract}
      \abstractText
      \newline
      \newline
    \end{abstract}
  \end{@twocolumnfalse}
]


\section{Introduction}
White Dwarfs represent one of the possible final stage of the stellar evolution. 

The study of compact object is an interesting problem as efforts from different fields are conveyed: high density environment requires both a quantum description and a relativistic one. 

\section{Equilibrium equations}
In this section the equations that regulate the hydro-static equilibrium of stars will be examined. Both the Newtonian case and the general relativistic case one will be discussed. 
\subsection{Newtonian equilibrium equations}
White Dwarf can be well-approximated as spherical; in the  Newtonian case, the request of equilibrium for a spherical distribution of mass implies the hydro-static equilibrium for an infinitesimal shell of radius $dr$ at $r$ from the center and leads to the following differential equation:
\begin{align}
\label{eq:stellarnewtonian}
    \begin{cases}
    \frac{dp}{dr} = -\frac{G m(r)\rho(r)}{r^2}\\
    \frac{dm}{dr} = 4\pi r^2\rho(r)
    \end{cases}
\end{align}
Where $m(r)$, $\rho$ and $p$ are mass, density and pressure at a given radius $r$. 

With the boundary conditions:
\begin{align}
\begin{split}
    &m(0) = 0\\
    &p(0) = p_c
    \end{split}
\end{align}
they define a system of non linear coupled differential equations. 

It is interesting to study the problem of uniqueness of solution of the differential problem.

\subsection{Uniqueness of the solution}
\label{sec:uniq}
In this section the problem of the uniqueness of the solution for the problem (\ref{eq:stellarnewtonian}) with given initial condition is discussed. 

As  K\"ahler, \& Weigert, appointed\cite{uniq} a real star evolve uniquely given its initial condition, the problem is whether the approximation we make yield to a unique solution.

K\"ahler showed\cite{uniq}\cite{uniqK} a condition for a local uniqueness in the static cases (the one we are treating). 

Even though for the general case the uniqueness cannot be guaranteed, by linearizing the equation uniqueness in the neighbourhood  of any equilibrium condition can be established.

The linear approximation is not truly limiting as in the pragmatic cases methods of solution for the numeric integration are based on the calculation of finite differences.

The 4 most general differential equations that govern stellar structure\cite{gleddening} can be expressed in function of pressure $p$, temperature $T$, radius $r$, Energy flux $L_r$ and $m(r)/M$ =  $\xi$, neglecting the time-dependence as  :
\begin{align}
& a = (p,T,r,L_r) &\\
\label{eq:dera}
    &\frac{d a_i}{d \xi} = f_i(\xi,a_1,a_2,a_3,a_4); & i=1, \cdots, 4;
\end{align}
Assuming $\nabla_{ad}$ and chemical composition given  $\epsilon_n = \epsilon_\nu = 0$ (nuclear  and neutrinos contributions to energy ). 

We can now solve for $\xi \in [\xi_\alpha , \xi_\beta]$ with boundary condition  at  $\xi_\beta$ near center and $\xi_\alpha$ near atmosphere imposing: 
\begin{align}
\label{eq:a1a}
    &a_1(\xi_\alpha) = \psi_1\left(a_3(\xi_\alpha), a_4(\xi_\alpha)\right)\\
\label{eq:a2a}
    &a_2(\xi_\alpha) = \psi_2\left(a_3(\xi_\alpha), a_4(\xi_\alpha)\right)
\end{align}
The conditions \ref{eq:a1a} \&   \ref{eq:a2a} are equivalent to fix the exterior boundary of the star (normally where $a_1 =a_2 = 0 $). 

Interior boundary condition expressed in function of some parameters $P_c$ and $T_c$ (central pressure and temperature) are need to be fixed in order to give meaning to the differential problem:
\begin{align}
&a_i(\xi_\beta) = \varphi_i (P_c,T_c) & i = 1,\cdots, 4
\end{align}

Since we are interested only in smooth solution for stellar structure, so we can safely assume that:
\begin{align}
    \begin{split}
        &\frac{\partial f_i}{\partial a_k} \in \mathcal{C}^0\\
        &\frac{\partial f_i}{\partial \xi} \in \mathcal{C}^0
    \end{split}
\end{align}

Note that, we have chosen $\xi_\beta$ \textit{near} the center as we can admit singularity at $\xi = 0$.

The problem with these hypothesis has a unique solution $a(\xi)$ determined by $a(\xi_\beta)$. The value of $a(\xi)$ on the outer boundary can therefore be seen as function of $P_c$ and $T_c$: $a_i(\xi_\alpha)=h_i(P_c,T_c)$.
Defining $g_i$ for $i = 1,2$:
\begin{align}
\label{eq:gi}
    g_i := h_i(P_c,T,c)- \psi_i\left(h_3(P_c,T_c), h_4(P_c,T_c)\right)
\end{align}
Imposing  that $g_i = 0$ implies that the problem of local uniqueness is equivalent to a unique solution of \ref{eq:gi} in terms of $(P_c, T_c)$ in a neighbourhood.

To discuss the neighbourhood uniqueness, two solutions $a(\xi)$ and $\bar{a}(\xi)$ that differ only for a small perturbation $\delta a$ can be considered.

Giving that, we can expand eq. \ref{eq:dera} linearly:
\begin{align}
    \frac{\partial \delta a}{\partial \xi} = \frac{\partial f_i}{\partial a_k}(\xi)\,\cdot\,\delta a
\end{align}

Then the two initial condition should differ only for a  $\begin{pmatrix}\delta P_c \\ \delta T_c \end{pmatrix}$ infinitesimal. 
Again we note that the general uniqueness cannot be proven and it has been demonstrated false in some numerical experiment, but a first-order uniqueness can be interesting as is it useful in applying the  finite differences method. 

With that aim we can expand \ref{eq:gi} to the first order:
\begin{align}
\label{eq:system}
    \begin{cases}
    \frac{\partial g_1}{\partial P_c} \delta P_c +  \frac{\partial g_i}{\partial T_c} \delta T_c = 0\\
    \frac{\partial g_2}{\partial P_c} \delta P_c +  \frac{\partial g_i}{\partial T_c} \delta T_c = 0\\
    \end{cases}
\end{align}
A more convenient way to express \ref{eq:system} is:
\begin{align}
\label{eq:rochè}
G \cdot \delta = \begin{pmatrix} \frac{\partial g_1}{\partial P_c} & \frac{\partial g_1}{\partial T_c}\\ \frac{\partial g_2}{\partial P_c} & \frac{\partial g_2}{\partial T_c}
\end{pmatrix} \cdot \begin{pmatrix}\delta P_c \\ \delta T_c \end{pmatrix} = 0 
\end{align}

Giving that \ref{eq:rochè} is a linear system, we can apply Rouché-Capelli theorem affirming that a solution is unique if:
\begin{align}
    \label{eq:sol}
\det[G] \neq 0 
\end{align}

Then we have derived that \ref{eq:sol} is a sufficient condition for local uniqueness in a linearized case. The argument can be further developed for time-dependent solutions\cite{uniqK} \cite{uniq}.

\subsection{General relativistic case}
In this section the Oppenhaimer-Volkoff equations will be derived. c=1 when not specified. 

In a static isotropic region of space-time, $g^{\mu \nu}$ is time independent. Chosing $x^\mu=\left(t,\theta, \phi\right)$ we have ,after imposing the spherical symmetry, that the line element is:
\begin{align}
\begin{split}
\label{eq:dtau}
    d\tau ^2 =& e^{2\nu(r)}dt^2 - e^{2\lambda(r)}dr^2 +\\&-  r^2(d\theta^2 + \sin^2\theta d\phi^2)
\end{split}
\end{align}

From which we can deduct both $g_{\mu \nu}$ and $R_{\mu \nu}$ \footnote{$R = \text{Tr} R_{\mu \nu}$}(The Ricci tensor).

The Eintein equations:
\begin{align}
\label{eq:einstein}
    &G^{\mu \nu} = R^{\mu \nu } - \frac{1}{2}g^{\mu \nu}R\\
    &G^{\mu \nu} = -8\pi G T^{\mu \nu}
\end{align}
Can be integrated, given $T^{\mu \nu}$. Since the spatial component of $u^{\mu}$ is zero (static assumption) we have that $T^{\mu \nu}$ is diagonal: $T^{0}_0 = \epsilon$ energy density and $T^{\mu}_{\mu} = -p$ ($\mu \neq 0$)the pressure given by the Equation of State(EoS). 

The (00) component of \ref{eq:einstein} gives:
\begin{align}
    e^{-2\lambda(r)} = 1 - \frac{8\pi G}{r}\int_0^r \epsilon(r)r^2 dr
\end{align}

In analogy with the Newtonian case we define:
\begin{align}
\label{eq:mass}
    M(r) = 4\pi \int_0^r \epsilon(r)r^2 dr
\end{align}
hence:
    \begin{align}
    \label{eq:struct1}
    \frac{dM}{dr} = 4\pi r^2 \epsilon(r)
\end{align}
represents the first equation for the stellar structure. 

In terms of $M(r)$ the other 3 Einstein equations can be deduced and, once the metric has been eliminated, we obtain\cite{gleddening} (G=c=1):
\begin{align}
\label{eq:einstein2}
\frac{dp}{dr} = - \frac{[p(r) + \epsilon(r)][M(r) + 4\pi r^3 p(r)]}{r [r-2M(r)]}
\end{align}

The equations \ref{eq:einstein2} are commonly known as \textit{Oppenheimer-Volkoff equations}, and represent the most important results of this section. They can be solved simultaneously with \ref{eq:mass} for the radial distribution of mass-energy density.

We can now re-write the differential problem \ref{eq:stellarnewtonian} generalised as ( G and c are now re-introduced):

\begin{multline}
\label{eq:struct2}
    \frac{dp}{dr} = - \frac{G m(r) \epsilon(r)}{c^2r^2}\bigg[1+\frac{p(r)}{\epsilon(r)}\bigg]\bigg[1+ \frac{4\pi r^3 p(r)}{m(r)c^2}\bigg]\\
    \times \bigg[1-\frac{2Gm(r)}{c^2r}\bigg]^{-1}
\end{multline}
\begin{align}
    \frac{dm}{dr} = \frac{4\pi}{c^2} r^2 \epsilon(r)
\end{align}
Eq \ref{eq:struct2} together with \ref{eq:struct1} and boundary conditions define the generalisation of the Newtonian problem \ref{eq:stellarnewtonian}
\subsection{Stability condition} 

\paragraph{Stability condition}
\label{par:stab}
In this section an important necessary condition for stability is discussed:
\begin{align}
    \frac{\partial M(\epsilon_c) }{\partial \epsilon_c} > 0 
\end{align}
The derivative of the total mass $M$ of the structure respect to the initial condition $\epsilon_c = \rho_C c^2$ should be always greater than zero in order to have a stable configuration. 

\begin{figure}
    \centering
    \includegraphics[width = \columnwidth]{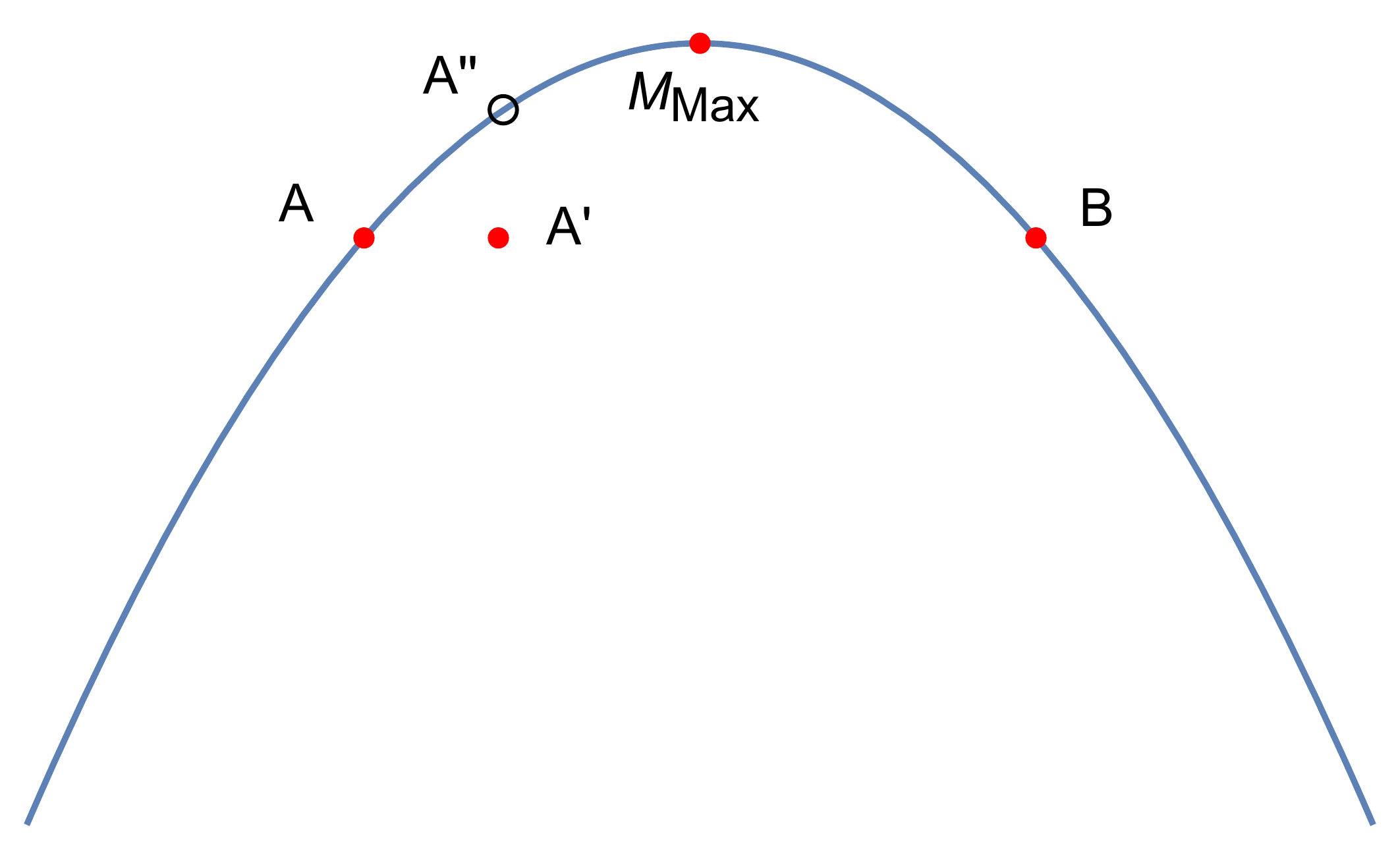}
    \caption{Situation described in text: neighbourhood of a maximum}
    \label{fig:graph}
\end{figure}

To prove the statement above, consider a neighbourhood of a stationary point in which the mass of the structure is a maximum, as displayed in Fig[\ref{fig:graph}]. In this case if the star is in the state A, where $\frac{dM}{d\rho}>0$ then a small perturbation that compresses the structure sends A in a point A'. The corresponding stable configuration A'' has a bigger mass so that the gravitational force pushes A' towards A; the same process happens whether A is perturbed with a small decompression. 

On the opposite, if B is perturbed, then  the new stable condition has a smaller mass, so that the perturbation is not re-adsorbed by gravitational attraction and the state is pushed away from B. \qedsymbol

The case in which:
\begin{align}
    \frac{\partial M(\epsilon_c) }{\partial \epsilon_c} = 0 
\end{align}
needs to be treated carefully: in this cases if $\Tilde{\epsilon_c}$ is a stationary points but $\forall \epsilon_c \in I(\Tilde{\epsilon_c}),  \frac{\partial M(\epsilon_c) }{\partial \epsilon_c} \geq0$ then the configuration is of equilibrium.
It is noticeable that the condition founded above is equivalent (in the temperature independent case) to the what established in the Newtonian case. This suggests that the analysis could be meaningful to repeat even in the relativistic case. 
\section{Equation of state}
In this section the Equation of State (EoS) will be discussed. The highly dense environment of stars requires a quantum treatment of matter, Beta-decay processes are also relevant in analysing the composition of stars. In this Chapter Coulomb interaction will be neglected, the discussion of the electric interaction will be further developed in sec \ref{sec:Coulombinteractions}. As we are interested in the static case, once the equilibrium has been reached, subtraction of energy by neutrinos will be also neglected even though they represent an important factor in the cooling-process of White Dwarf. 

\subsection{Fermi-gas equation of state}
As nucleons and electrons are all fermions they obey to the Fermi-Dirac statistic. 

If $n$ is the numerical density, the quantum properties of gases cannot be neglected when
\begin{align}
\chi := \frac{\Lambda}{a} \gtrsim 1
\end{align}
Where $\Lambda$ is the De Broglie thermal wavelength and $a = \left(\frac{4}{3}\pi n \right)^{-1/3}$. In a typical core of White Dwarf for electrons $\chi \simeq 96 $, so that the gas can be considered as highly degenerate. In a first-order approximation, the gas can be considered completely degenerate and treated at zero temperature. 

In addiction to this, for electrons relativistic endeavours should be considered when:
\begin{align}
    X_r := \frac{p_F}{m_ec^2} \gtrsim 1
\end{align}
Where $p_F$ is the Fermi-momentum. In typical White Dwarf condition $X_r$ values more than one. For a density $\rho = \num{2e6}$g/cm$^3$, $X_r \sim 1$. The transition to the relativistic region is responsible for the presence of a limiting mass in the Chandrasekhar model. 

\subsection{Completely degenerate equation of state}
In this section the EoS for a completely degenerate Fermi-gas is computed. 

We therefore compute the limit $T \to 0$ of the Fermi Distribution function over quantum states given that the degeneracy of each state is g = 2s+1 = 2 (spin = 1/2).

The Fermi Distribution over state is\cite{landau}:
\begin{align}
    n_p = \frac{1}{e^{{\beta(\epsilon - \mu)}}+1}
\end{align}
Where $\beta=\frac{1}{kT}$, $\epsilon$ is the energy of the level and $\mu$ is the chemical potential. 

\begin{align}
    \lim_{T\to0} n_p(T) = \Theta(\epsilon - \epsilon_F)
\end{align}
Where $ \epsilon_F:= \lim_{T\to0} \mu(T)$

We can therefore obtain the distribution over momentum requiring that in the completely degenerate case all the levels are fully occupied and the highest one have $p = p_F$
\begin{align}
\label{eq:n}
N/V = n = 2\cdot\frac{4\pi}{h^3}\int_0^{p_F} p^2 dp = \frac{p_F^3}{3\pi^2\hbar^3}
\end{align}
Hence the distribution function over momentum is:
\begin{align}
    f(p) = \begin{cases}
    \frac{4\pi}{\pi^2 \hbar^3}p^2 & p\leq p_F\\
    0                           & p>p_F
    \end{cases}
\end{align}
Then Energy  $\mathcal{E}=  E/V$ and the pressure are now computable:
\begin{align}
    &\mathcal{E}= \int_0^{p_F} f(p)\sqrt{(mc^2)^2 + (pc)^2} dp\\
    &P = \frac{8\pi c}{3h^3}\int_0^{p_F}p^3 \frac{p}{\sqrt{(mc^2)^2 + (pc)^2}}dp
\end{align}
The calculation has been carried out in the Appendix \ref{sec:integrals}, the results are:
\begin{multline}
\label{int:E}
    \mathcal{E} = \frac{c}{8\pi^2\hbar^3}\Big\{p_F\left(2p_F^2 + m^2c^2\right)\sqrt{p_F^2 + m^2c^2} \\ -(mc)^4\arcsinh{\frac{p_F}{mc}}\Big\}
\end{multline}

\begin{multline}
\label{int:P}
    P = \frac{c}{8\pi^2\hbar^3}\Big\{p_F\left(\frac{2}{3}p_F^2 - m^2c^2\right)\\ \times \sqrt{p_F^2 + m^2c^2} +(mc)^4\arcsinh{\frac{p_F}{mc}}\Big\}
\end{multline}
\paragraph{Non-relativistic limit}
In the non-relativistic limit the EoS has a simple form. Expressing $p_F$ in terms on n \ref{eq:n}, when $p_F\ll mc$:
\begin{align}
    &P = \frac{(3\pi^2)^{2/3}}{5} \frac{\hbar}{m}n^{5/3}\\
    &\mathcal{E} = \frac{3\hbar^2}{10}\frac{3\pi^2}{m}n^{2/3}
\end{align}

\paragraph{Ultra-relativistic limit}
In the ultra-relativistic case, when $p_F \gg mc$ we have:
\begin{align}
    &\mathcal{E} = \frac{3\hbar c}{4}(3\pi^2)^{1/3}n^{4/3}\\
    & P = \frac{\mathcal{E}}{3}
\end{align}

\section{Chandrasekhar approach}
We are now ready to compute the mass-radius relation in the Chandrasekhar limit for a White Dwarf. 

In lack of nuclear reaction, the main contribution to the pressure is given by the degeneracy pressure of electrons, so that we can approximate all the pressure with the pressure of the Fermi-Dirac electrons gas. The nucleons pressure is negligible as in $\chi_n \simeq \chi_e/2000$   

\subsection{Polytropes method}
An equation of state in which $P=P(n) = K n^\gamma$ is called a polytropic equation of state(No temperature dependence).  We note that for a ultra-rel and non-rel degenerate gases have polytropic EoS: $P_{nr} = K n^{5/3}$ and $P_{ur}=K n^{4/3}$. 
Polytropes are very powerful methods to find a semi-analytical solution to the differential problem \ref{eq:stellarnewtonian}.

In the Newtonian approximation when a polytropic relation with index $n = \frac{1}{1-\gamma}$ is assumed $P = K\rho^\gamma$ ($\rho$ is the mass-density) , the equations \ref{eq:stellarnewtonian} can be reduced to the Lane-Emden equation:
\begin{align}
    \frac{1}{\xi^2}\frac{d}{d\xi}\,\xi^2\,\frac{d\phi}{d\xi} + \phi^{n} = 0 
\end{align}
In terms of a-dimensional radius and density:
\begin{align}
&\xi = \frac{r}{\lambda} &\lambda = \Big(\frac{(n+1)K\rho_c^{\frac{1-n}{n}}}{4\pi G}\Big)^{1/2} \\
& \rho = \rho_c \phi^n
\end{align}
Where $\rho_c$ the parameter that fixes the central density of the star. 

Two initial conditions are required as the first-order differential system \ref{eq:stellarnewtonian} has been transformed in a second-order ODE:
\begin{align}
    &\phi(\xi = 0) = 1 \\
    \label{eq:maxden}
    &\frac{d\phi}{d\xi}(\xi = 0) = 0
\end{align}
The condition \ref{eq:maxden} impose that the density is maximum at the center.

Once the ODE is solved for the a-dimensional density $\phi(\xi)$ we can obtain the radius of the structure $\xi_*$ as the smallest zero of the solution:
\begin{align}
\begin{split}
   &\xi_* = \min\limits_{\xi_i}\{\xi_i \,|\,\phi(\xi_i)=0\}
   \end{split}
\end{align}
Therefore the total mass M is:
\begin{align}
    &M(R) = 4\pi \int_0^R r^2 \rho(r)dr =
    \\
        \label{eq:prelam}&=4\pi\rho_c\lambda^3\int_0^{\xi_*}\phi^n(\xi)\xi^2 d\xi=\\
    \label{eq:postlam}
    &= -4\pi\rho_c\lambda^3\bigg[\xi^2 \frac{d\phi}{d\xi}\bigg]_{\xi_*}
\end{align}
Note that we have used the Lane-Emden equation from \ref{eq:prelam} to \ref{eq:postlam}.

With the aim of finding the limit mass for a White Dwarf, we are mainly interested in the asymptotic case in which electrons have an ultra-rel EoS : $\gamma = 4/3 \Rightarrow n=3$.
\subsection{Solutions to Lane-Emden equation}
The Lane-Emden equation admits analytical solutions only for n = 0,1,5; in the other cases it has to be integrated numerically. 
\begin{figure}
    \centering
    \includegraphics[width = 0.5\textwidth]{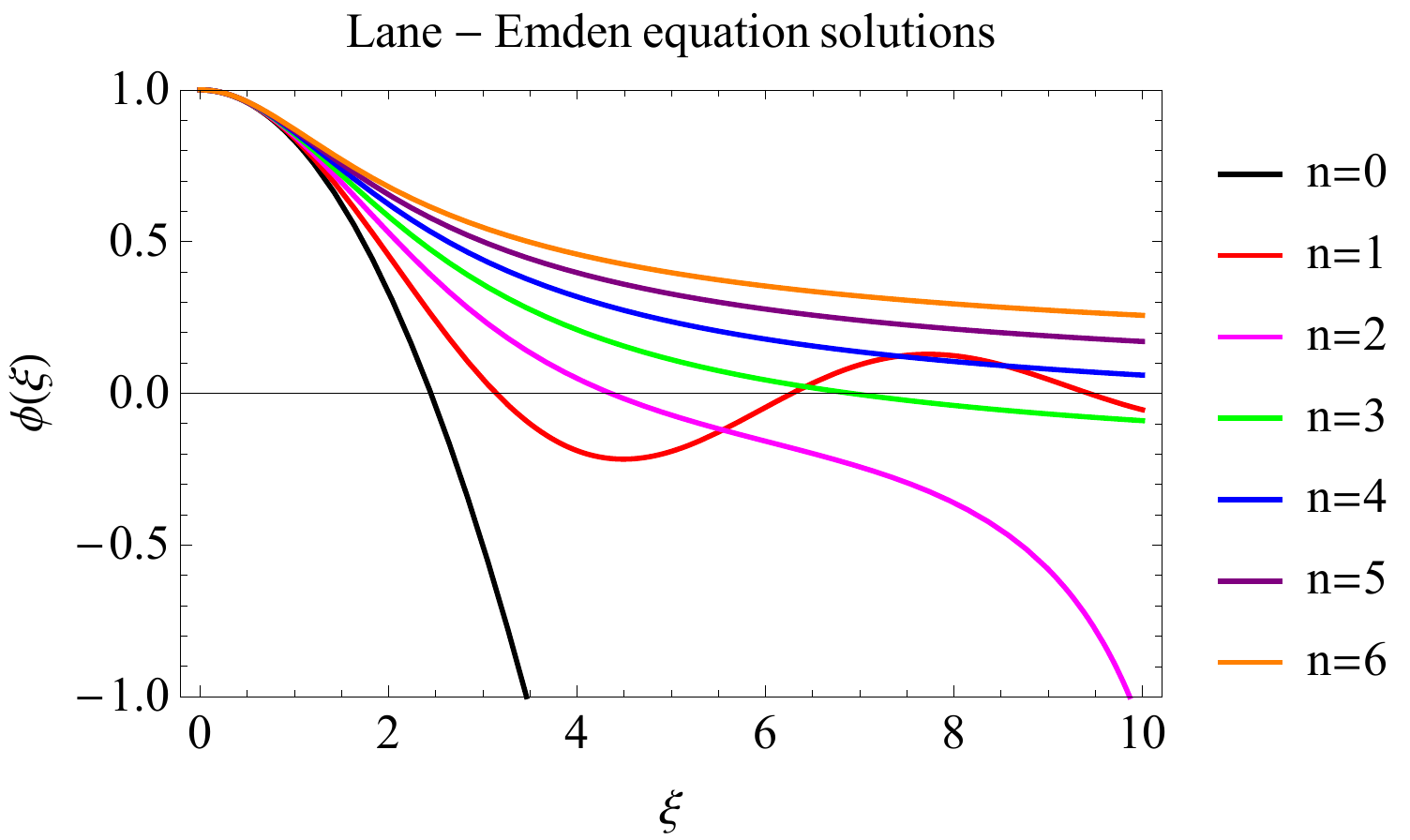}
    \caption{Numerical solutions for The Lane-Emden equation for seven values of index}
    \label{fig:LaneEmden}
\end{figure}
In Fig[\ref{fig:LaneEmden}]  solutions to the equation have been evaluated numerically and analytically (when possible) for seven different values of the polytropic index using 
\textit{Mathematica}.

For index $n>0$  polytropes has an infinite extension. 
\subsection{The Chandrasekhar mass}
In this section the Chandrasekhar mass value is derived. 

For n = 3  the numerical solutions for the relevant quantities are:
\begin{align}
&\xi_*&&=6.89685 \\ \bigg[\xi^2 &\frac{d\phi}{d\xi}\bigg]_{\xi_*} &&= -2.01824
\end{align}
Given that ($n = \frac{\rho}{\mu_e m_H}$where $\mu_e = 2$\footnote{The mean electronic weight in the case of \ce{He}, \ce{C} and \ce{O}} is the overall electronic weight and $m_H$ is the Hydrogen mass)
\begin{align}
    &\lambda = \bigg[\frac{K }{\pi G}\bigg]^{1/2} \rho_c^{-1/3}\\
    & K =  \frac{3}{\pi}^{1/3}\frac{h c}{8 \left(2m_H\right)^{4/3}}
\end{align}

We note that in the special case of $n = 3$ the total mass of the polytropos do not depend on the central density, we can therefore state that it represent a limiting mass for the structure, known as the Chandrasekhar mass:
\begin{align}
    M_{Ch} = 1.43469 M_{\odot}
\end{align}
Where $M_{\odot} = \num{1.989 e33}\,$g is the Sun mass.

\subsection{Solution with complete EoS}
In this section the problem is treated in a more general way, using the complete EoS for a Fermi-gas. 

We need to solve the differential problem \ref{eq:stellarnewtonian}
with the complete EoS:
\begin{align}
    &P = \mathcal{K}\bigg[(2\xi^3-3\xi) \sqrt{1+\xi^2} + 3 \arcsinh^{-1}\xi\bigg]\\
    &\mathcal{K}=\frac{(m c)^4 \pi c}{3 h^3}, \hspace{6mm} \xi = \frac{p_F}{m c}
\end{align}

We have that the baryon density, in terms of $\xi$ is:
\begin{align}
    \rho = \frac{(mc)^3\mu_e m_H(g)}{3\pi^2\hbar^3}\xi^3 := \chi \xi^3 
\end{align}
Introducing the variables: $\mu =m/M_\odot$ and $x = r/R_\odot$, the differential problem has the form: 
\begin{align}
    &\frac{dm}{dr} =4\pi \chi\xi^3 r^2\\
    &\frac{d\xi}{dr}=-\left(\frac{G\chi}{8\mathcal{K}}\right)\sqrt{1+\xi^2}\frac{ m}{r^2 \xi}
\end{align}

This equations can be simultaneously solved for $m(r)$ and $\xi(r)$ given the initial conditions:
\begin{align}
    &\xi(0) = \frac{1}{\mathcal{K}}\rho_c^{1/3}\\
    &m(0) = 0
\end{align}

This differential system can be numerically integrated, in that case I used NDSolve of \textit{Mathematica}, it uses the Runge-Kutta based on finding the coefficient for the Taylor Series associated to the solutions. I used also Numpy methods on Python but since the algorithm in that case is not released I report the results obtained with \textit{Mathematica}.

Once the solution has been founded, the Mass of the structure can be determined by looking for the point in which $\xi[r]$ change sign. In fact, the zero of the density represent the external limit of the structure (cfr. section for boundary conditions \ref{sec:uniq}). For numerical purpose of stability the initial conditions cannot be imposed ad $r=0$, instead it is more convenient to fix it on the \$MachineEpsilon\footnote{Gives the difference between 1 and the next smallest number representable \$MachineEpsilon = \num{2.22045e-16}}.

\begin{figure}
    \centering
    \includegraphics[width = \columnwidth]{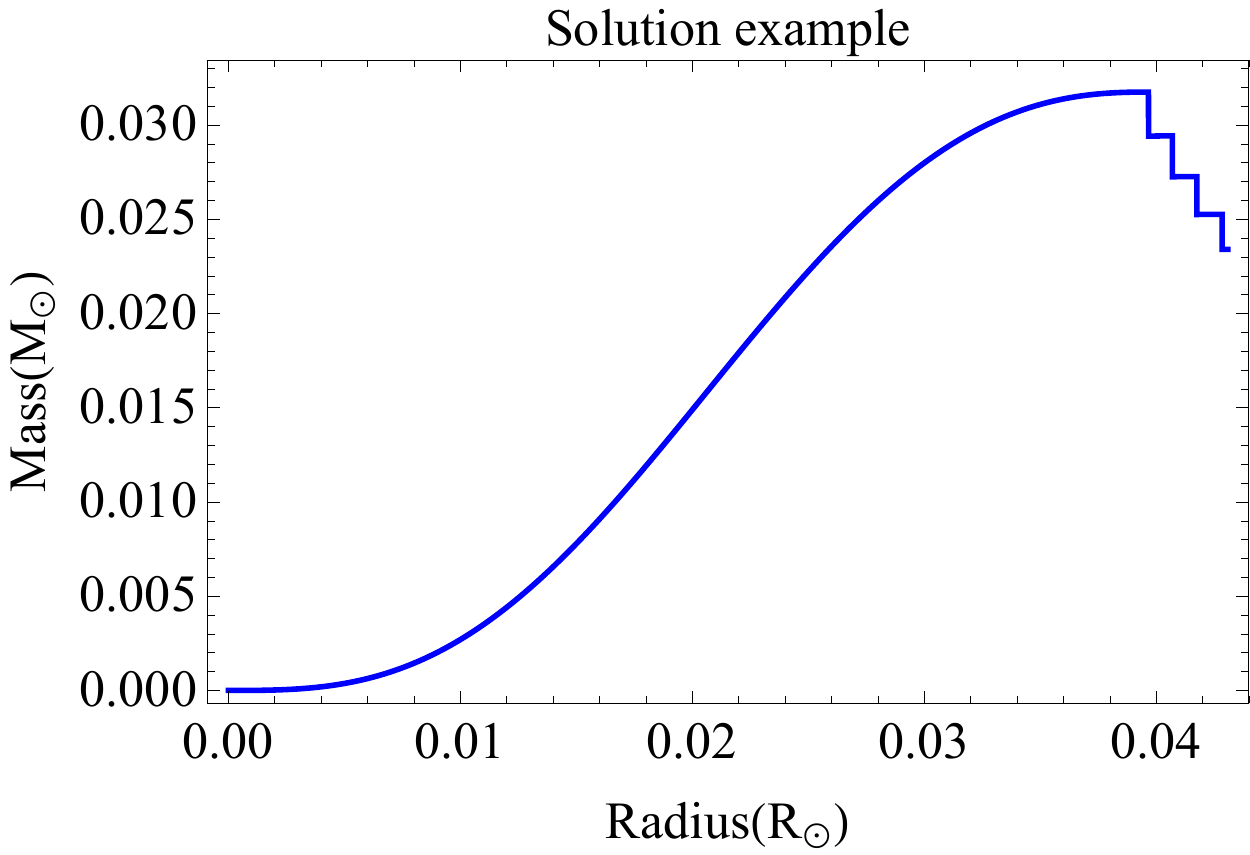}
    \caption{Solution for $\rho_c = 10^{6.7}$}
    \label{fig:100sol}
\end{figure}

In Fig [\ref{fig:100sol}] an example of the solution, it is clear from the plot the limit of the structure, we can see how after the change in sign of the density the mass decrease: a non-physical situation. 

\begin{figure}
    \centering
    \includegraphics[width = \columnwidth]{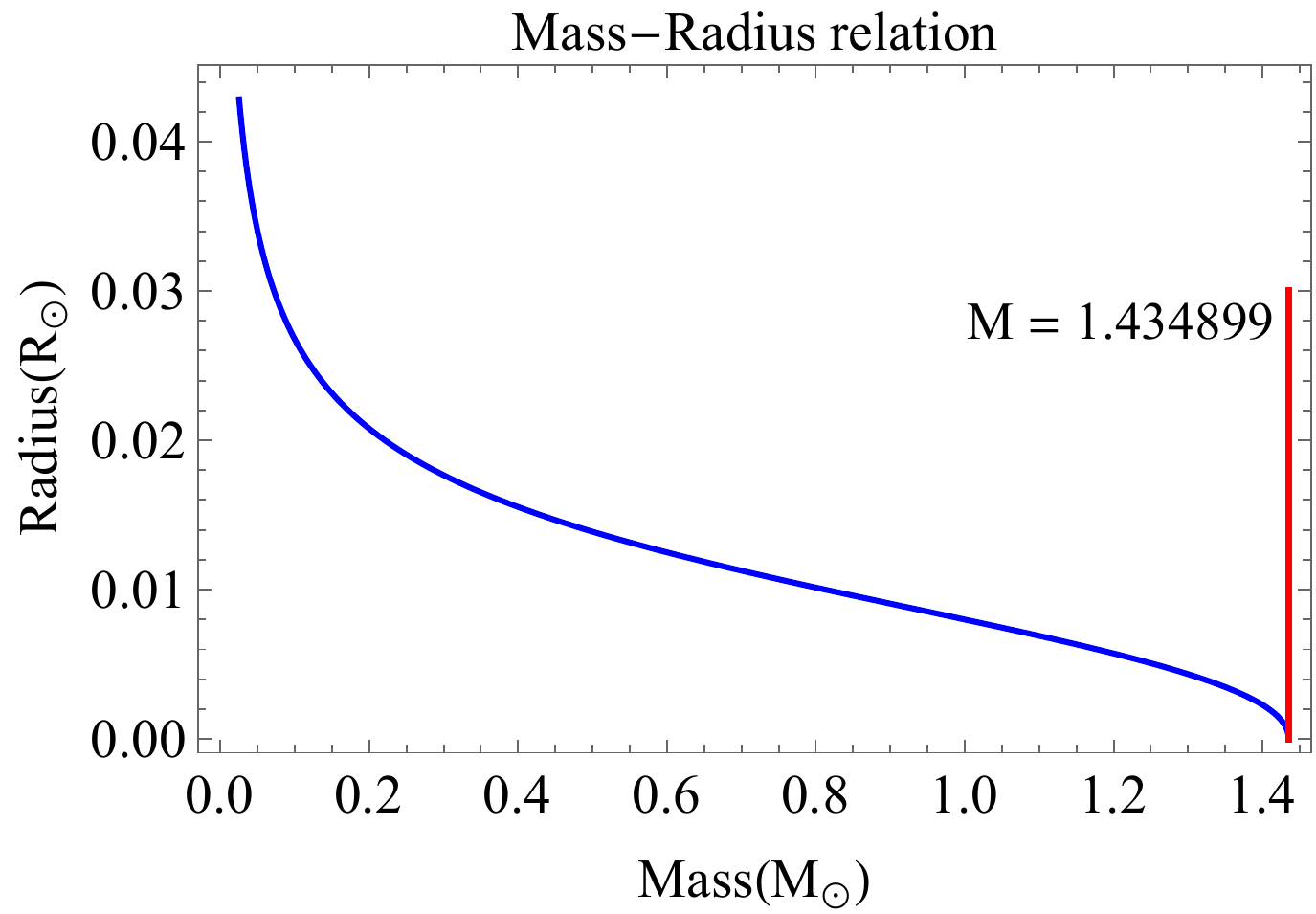}
    \caption{Mass-Radius relationship for a White Dwarf}
    \label{fig:massradius}
\end{figure}

The procedure has been repeated extracting the couples $(M_\text{Max},R_{\text{Max}})$ and $(M_\text{Max},\rho_C^{\text{Max}})$ from the solution, varying the central density. In Fig[\ref{fig:massradius}] the results obtained.
As expected in the limit for which $\rho_c \to \infty$ the Chandrasekhar limiting cases is valid and the Mass tends to a vertical asymptote such that the configuration is stable. The numerical value of the latter is $M = 1.434899$ that differs only for $0.015\%$ with $M_{Ch}$ founded solving Lane-Emden equation in the ultra-relativistic limit. 

\begin{figure}
    \centering
    \includegraphics[width = \columnwidth]{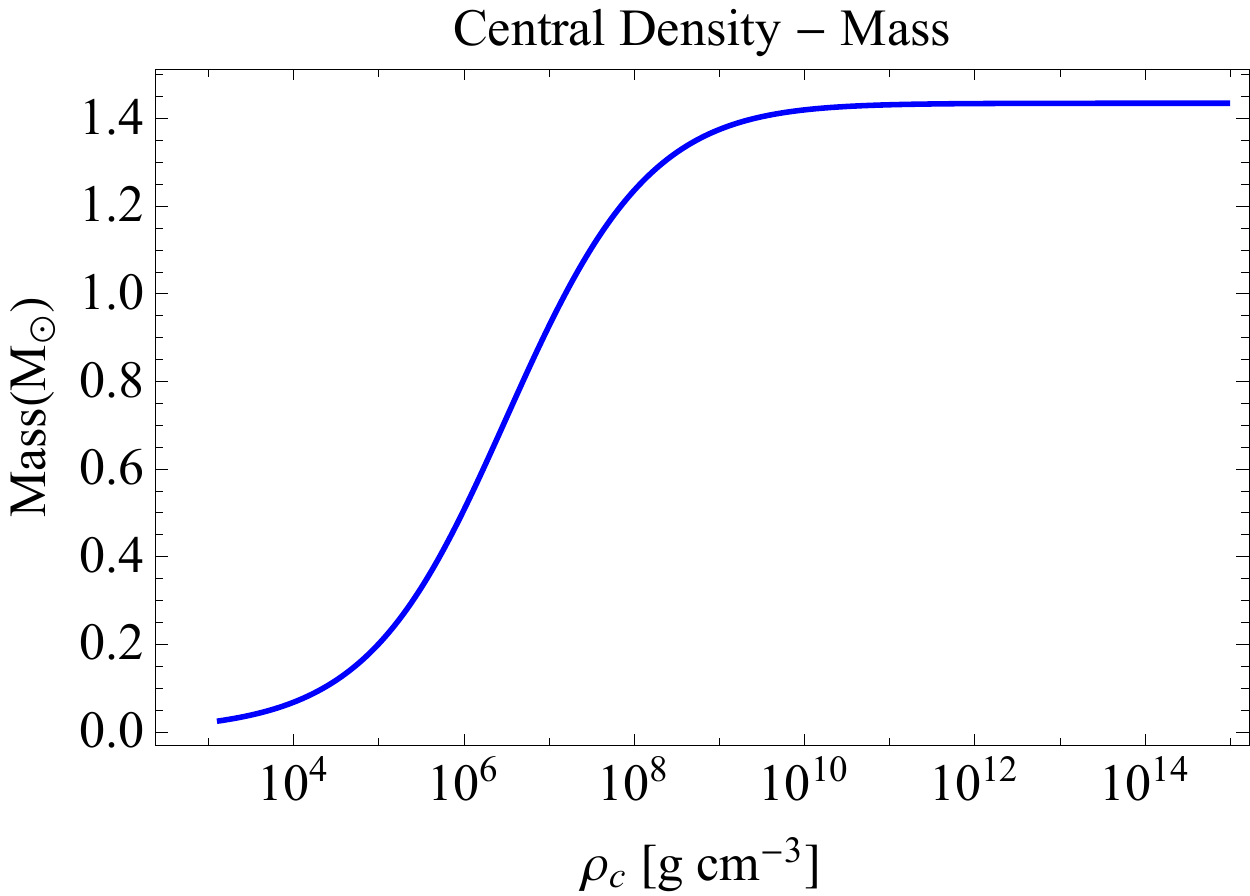}
    \caption{Mass-Central density}
    \label{fig:pressuremass}
\end{figure}

An important feature is that, in this model, all the states obey the condition necessary for stability proven in \ref{par:stab}. In Fig[\ref{fig:pressuremass}] the results. 

\section{General relativistic corrections}
In the complete Chandrasekhar solution, there is no lower bound to the radius as it decreases when the central density increases so that the total mass remains the fixed $M_{Ch}$. If we do not consider other effects, this remains true whether the radius of the structure is big compared to the Schwarzschild radius of the body i.e.:
\begin{align}
    R > \frac{2GM_{Ch}}{c^2}\simeq &\num{6.0855e-06} \, R_\odot \\=&  \num{4.2376}\,\,\text{km}
\end{align}
For $R\sim \frac{R_\odot}{500}$ (i.e. $\rho\gtrsim \num{1e10}\,\,\text{g cm$^{-3}$}$ relativistic effects cannot be ignored anymore.

With the aim of consider relativistic corrections we integrate the structure equations with a approximation $\epsilon(r) \simeq \rho(r)c^2$, where $\rho(r)$ is the baryonic mass density. 

This approximation is valid because:
\begin{itemize}
    \item The baryonic matter momentum  $|p_b|^2 = 0$  in the static case;
    \item Even though electrons have $p_F \gg m_e$ in the worst situation considered in the non relativistic approximation $\rho_c = \num{5e14}$ $p_F = 981 m_ec\simeq \SI{490}{\mega\eV\per c}$ and became less than $400 m_ec$ at $R = R_{\text{TOT}}/1000$, so that in the most of the cases $p_F c\ll m_p c^2$;
            \item For Baryons $\frac{p_F}{m_p c}\simeq \xi/2000$ therefore Fermi  momentum of Baryons is also negligible.
\end{itemize}

The following differential problem can be solved numerically with the same boundary conditions used before:
\begin{align}
    \frac{dm}{dr} = \frac{4\pi}{c^2}\epsilon(r)
\end{align}
\begin{multline}
    \frac{d \xi}{dr} = -\frac{d\xi}{dp}\frac{G m(r) \epsilon(r)}{c^2r^2}\bigg[1+\frac{p(r)}{\epsilon(r)}\bigg]\\\times\bigg[1+ \frac{4\pi r^3 p(r)}{m(r)c^2}\bigg] \bigg[1-\frac{2Gm(r)}{c^2r}\bigg]^{-1}
\end{multline}
Where $\epsilon(r) = \mathcal{K}c^2 \xi^3$ and $p= p(\xi)$ the complete EoS for a degenerate electrons gas \ref{int:P}.

\begin{figure}
    \centering
    \includegraphics[width = \columnwidth]{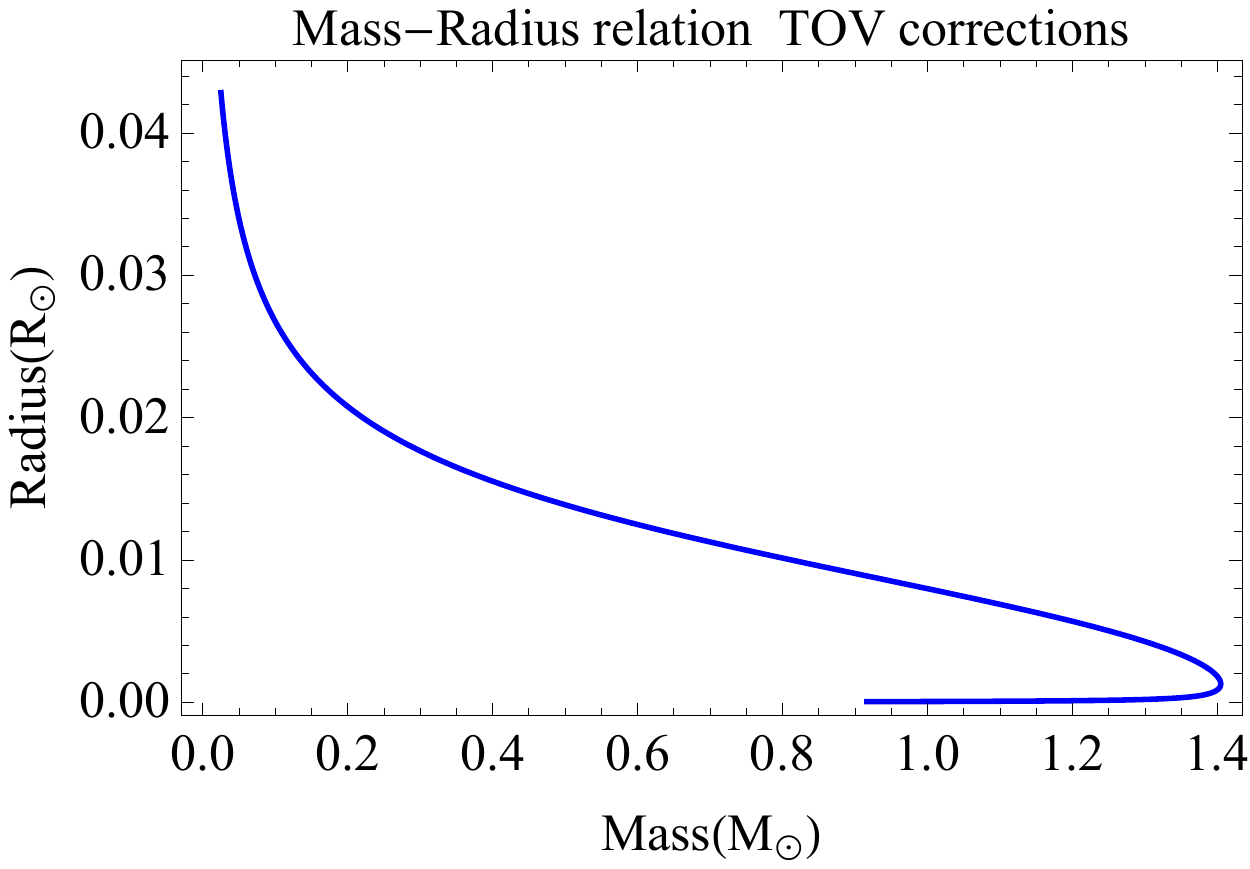}
    \caption{Mass Radius relation with correction due to the relativistic metric }
    \label{fig:tovmass}
\end{figure}

In Fig[\ref{fig:tovmass}]
In this level of approximation there is no presence of a Chandrasekhar mass.
Nevertheless, there is a maximum mass at 
\begin{align}
&M_\text{Max}= 1.404565\,M_\odot &
\end{align}
In correspondence to the maximum mass, the radius and the central pressure of the structure is:
\begin{align}
    &R\left(M_\text{Max}\right) &&= 0.001280\,R_\odot =\nonumber \\
& && = 0.1399 R_{\text{Earth}}\\
&\rho_c(M_{\text{Max}}) &&=  \num{1.7783e10} \SI{}{g\per cm^3}
\end{align}
This value can be considered as the maximum mass for a White Dwarf. 
\begin{figure}
    \centering
    \includegraphics[width = \columnwidth]{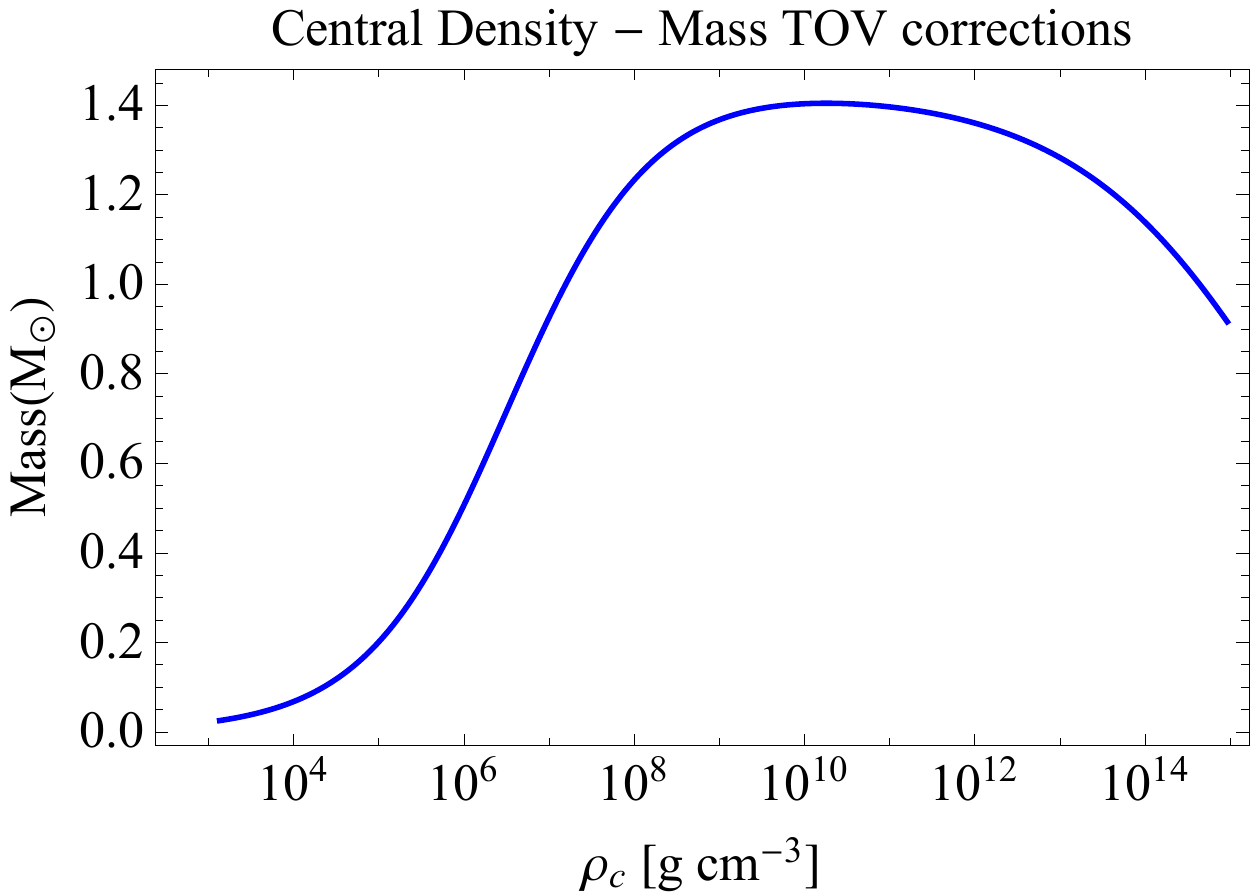}
    \caption{Mass - Central density}
    \label{fig:tovcentral}
\end{figure}
Interesting information can be deduced from the relation between mass and central density. In Fig[\ref{fig:tovcentral}] is plotted the total mass in function of the central density $\propto \epsilon_c$. Because of the stability conditions (cfr. \ref{par:stab}) $M_\text{Max}$ is also a limiting mass as beyond that all the configurations are unstable. This limiting mass differ from $M_{Ch}$ by $2.1\%$.

Both in the Chandrasekhar and in the Relativist model with complete EoS there is no dependence from the temperature and, from the chemical composition, therefore they represent simplifications of the problem. Nevertheless, the results obtained describe the correct magnitude of the mass-radius relation for the stars.

\section{Improvements to the Equation of state}

Matter in white dwarfs is not at its lowest energy state: neutrons and protons are bounded in nuclei that are not in their lowest state possible ($\ce{^{62}Ni}$ or $\ce{^{56}Fe}$, depending on whether the total binding energy or the binding energy per nucleons is considered) as in the neutron star case. Because of that, it is possible to extract energy further energy trough beta-decay and pico-nuclear reactions(The latter will not be analysed in this paper)
\subsection{Charge neutrality}
As White Dwarf are mainly constituted by ionized atoms and free electrons, it is important to study the electric properties of matter. 

The limit for which particle with charge $e$ and mass $\mu$ will not be expelled from the external surface of the star by Coulomb Force is given by:
\begin{align}
    \frac{Z_{\text{tot}}e^2}{R^2} < \frac{GM\mu}{R^2}
\end{align}
In the general relativistic case $M > Am_p$ (where A is the total number of baryons of the star). In the worst case $\mu = Zm_p$ Z $\leq 16$ (Oxygen White Dwarf) so that:
\begin{align}
   \frac{ Z_{tot}}{A Z} < \left(\frac{m}{e}\right)^2 \simeq \num{e-36}
\end{align}
So that the mean electric charge per nucleons is $\sim Z \num{e-36}$: White Dwarf can be considered electrically neuter. 

\subsection{$\beta$-equilibrium}
\label{sec:beta}
As the mass of neutron is greater than the sum of proton and electron by:
\begin{equation}
    \Delta :=  (m_n -m_p -m_e)c^2 
\end{equation}
in normal condition inverse $\beta$ decay are forbidden.
\begin{align}
    p + e^- \rightarrow n + \nu_e
\end{align}
But in a degenerate environment whenever happens that:
\begin{align}
    E_F := E - m_e c^2 \geq \Delta 
\end{align}
Is more convenient energetically for an electron to decade $\beta$ rather than be in a phase space cell with $P_F$. 

White Dwarf equation of state should be corrected. 

If we consider a 3 component plasma (simplest situation possible) made of free protons, neutrons and electrons, we need to minimise for every configuration the total energy of the plasma:
\begin{align}
\label{eq:energymin}
\epsilon(n_n, n_p, n_e) = \epsilon(n_n) + \epsilon(n_p) + \epsilon(n_e)
\end{align}
Where $\epsilon(n_i)$
We have two different constraints:
\begin{align}
    &n_n + n_p = \text{const.} := n_b \\
    & n_p = n_e
\end{align}

The first one is the expression of the conservation of Baryion number and the second is the overall neutrality of charge. 
Since the three particles are all fermions we can use eq \ref{int:E} for $\epsilon(n)$

To minimise \ref{eq:energymin} we can use the Lagrange multipliers method, we define:
\begin{equation}
    F := \epsilon(n, n_p, n_e) + \alpha (n_p + n_n - n_b) + \beta (n_p - n_e)
\end{equation}
We need impose:
\begin{align}
\label{eq:derivative}
    \frac{\partial F}{\partial n_i} = 0 
\end{align}
A clever way to calculate \ref{eq:derivative} is using the integral expression for $\epsilon$:
\begin{align}
    &\frac{\partial \epsilon}{\partial n} = \frac{\partial \epsilon}{\partial p} \frac{\partial p}{\partial n} = \\&= \frac{1}{8\pi^2 \hbar^3} \sqrt{(mc^2)^2 + (pc)^2}p^2 \cdot \frac{8\pi^2 \hbar^3}{p^2}\nonumber=\\& =   \sqrt{(mc^2)^2 + (pc)^2}
\end{align}
Imposing \ref{eq:derivative} and eliminating the Lagrange multipliers we obtain:
\begin{multline}
\label{eq:chemical}
    \sqrt{p_p^2 + (m_pc)^2} + \sqrt{p_e^2 + (m_ec)^2}  =\\ =\sqrt{p_n^2 + (m_nc)^2} 
\end{multline}
We can recognise in \ref{eq:chemical} the general condition on the chemical potential at equilibrium when coexist different phases:
\begin{align}
\label{eq:mui}
    \mu_e + \mu_p = \mu_n
\end{align}
Eq. \ref{eq:mui} ensures that all the levels below the Fermi level are filled and that no energy can be further extracted from the gas by $\beta$ decay. Clearly this is only an approximation as for low density the configuration in which $n_n = 0$ is the lowest possible. 

The procedure above can be easily generalised \cite{landau}: in general we have $\beta$ decay:
\begin{align}
\ce{^n_ZX} +e^- \rightarrow \ce{^{n+1}_{Z-1}Y}
\end{align}
that can be written in compact form as $\sum_i \nu_iX_i = 0$. If the decay happens at constant temperature and pressure then the potential $\Phi(\nu_i)$ must be minimum given P and T. If we impose that the total derivative of $\Phi$ respect to the concentrations $n_i$ is zero we obtain:
\begin{align}
    \sum_i \nu_i \frac{\partial \Phi}{\partial n_i} = 0
\end{align}
Or:
\begin{align}
    & \mu_i = \frac{\partial \Phi}{\partial n_i}\\
    \label{eq:general}
     &\sum_i \nu_i \mu_i = 0
\end{align}
Eq. \ref{eq:general} is \ref{eq:mui} in the special case of a three component plasma. 

Using the relation eq. \ref{eq:n} between n and p, we can obtain the chemical composition point for point.
Do to that we integrate $\forall n$ fixed :

\begin{align}
\begin{cases}
\frac{1}{3\pi^2\hbar^3} (p^3_n + p^3_p) = n\\
\mu_e(p_e) + \mu_p(p_p) = \mu_n(p_n)\\
\end{cases}
\end{align}
Together with $p_p = p_e$ (charge neutrality) and with the Fermi EoS for p. 

Hence:
\begin{align}
    p_n= \bigg(3\pi^2 \hbar^3 n - p^3_p \bigg)^{1/3}
\end{align}

I have numerically integrated the equations above. 
It is convenient to use the a-dimensional variable $\xi = \frac{p_F}{m_ec}$.

\begin{figure}
    \centering
    \includegraphics[width = \columnwidth]{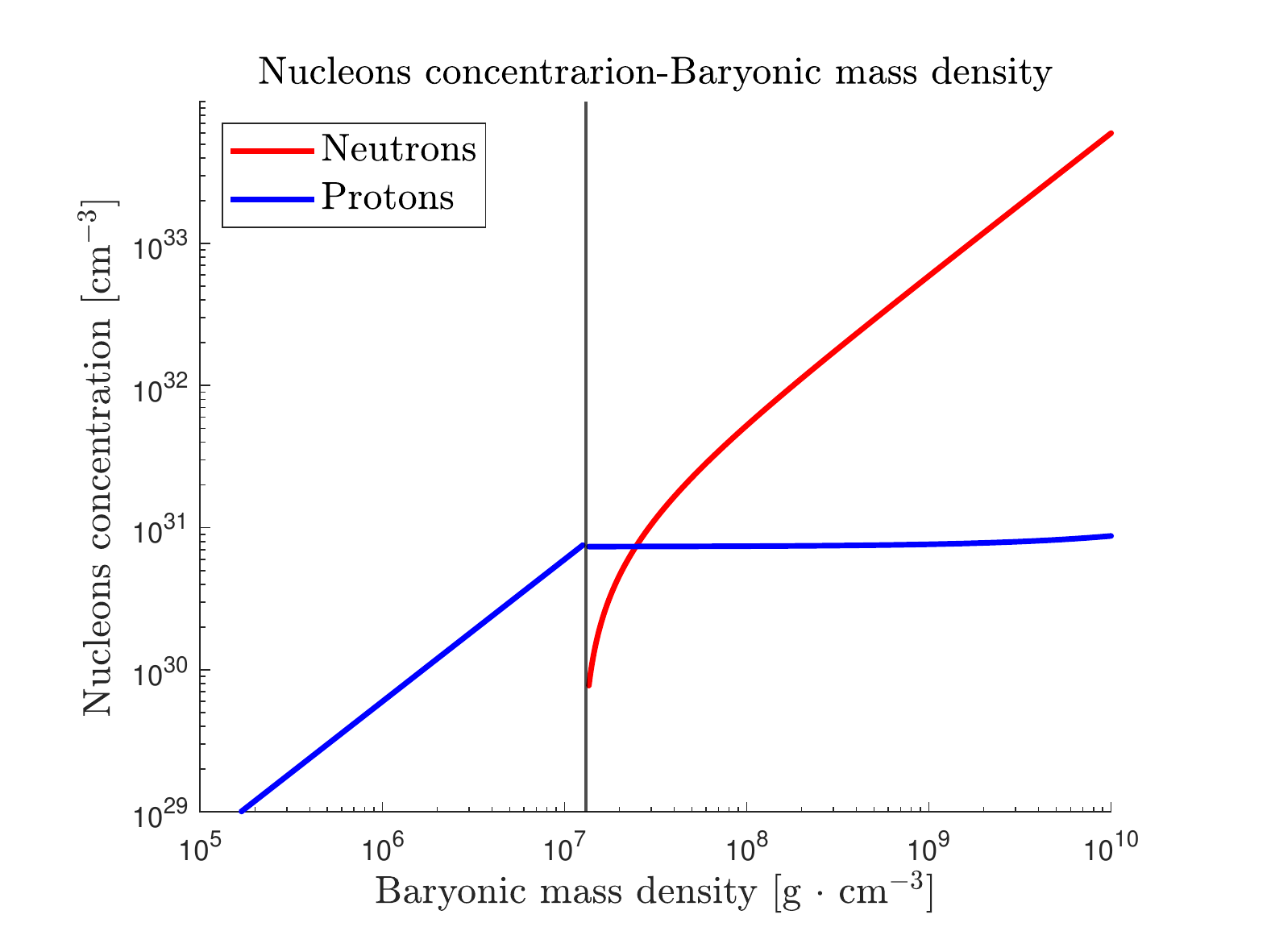}
    \caption{Protons and neutrons concentration at $\beta$ equilibrium}
    \label{fig:conc}
\end{figure}

\begin{figure}
    \centering
    \includegraphics[width = \columnwidth]{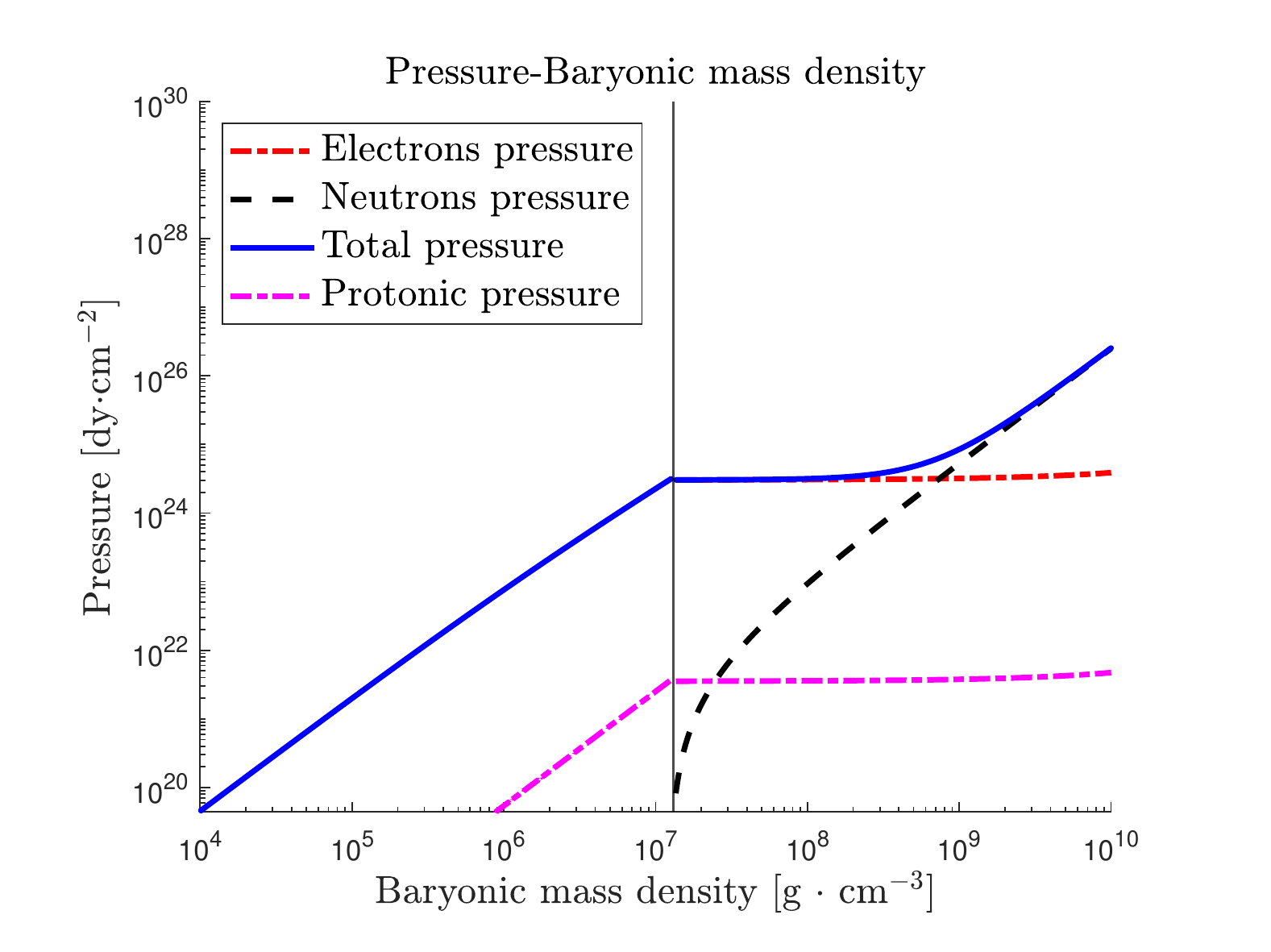}
    \caption{Pressure of the mixture at $\beta$ equilibrium}
    \label{fig:totalpressure}
\end{figure}

In Fig[\ref{fig:conc}] and Fig[\ref{fig:totalpressure}] the results obtained

The numerical threshold for $\beta$ decay is $\bar{\rho} = \SI{2.37e7}{\g\per\cm^{3}}$.
The estimation with $\Delta$ leads to a value of $\bar{\rho} = \SI{1.2e7}{\g\per\cm^{3}}$ almost an half of what founded. This is mostly because in the numerical calculation we have considered the fact that neutrons are themselves fermions. 

As expected for $\rho < \bar{\rho}$ pressure is dominated by electrons and a compression lead to an increase in concentration of protons and electrons.

Once $\rho>\bar{\rho}$\, ,\, $p+ e^- \rightarrow n + \nu$ can happen: a further compression do not results in an increase of pressure, instead it increases the concentration of neutrons in the mixture subtracting degenerate electrons. Asymptotically all protons and electrons  are converted in neutrons who became the only source of pressure. 

\section{Stellar structure at $\beta$ equilibrium}
$\beta$-decay are responsible for a source of instability in WD. Once the threshold density for $\beta$-decay is reached stars became unstable: as electrons are subtracted from the plasma the total pressure of electrons no longer balances the gravitational force. 

Since we are interested in computing the mass-radius relation for White Dwarf considering $\beta$ equilibrium, the chemical composition can no longer be neglected.

To obtain an estimation of the effect, we can make the simplification for which, the star is considered a three-component plasma only \textbf{after} the threshold is reached

\begin{figure}
    \centering
    \includegraphics[width = \columnwidth]{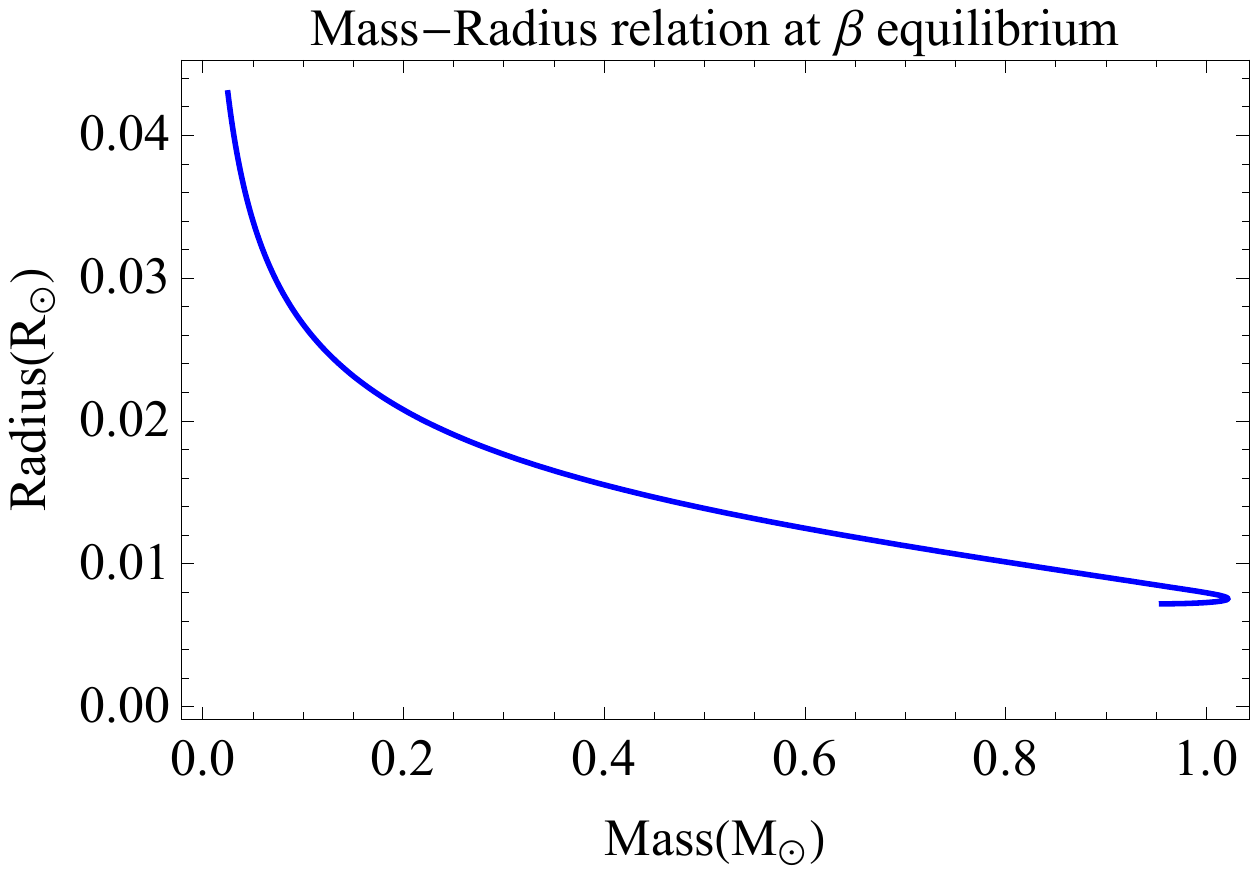}
    \caption{Mass radius relation, TOV equations @ $\beta$ equilibrium $p,n,e$ mixture}
    \label{fig:tovbeta}
\end{figure}

The modified equation of state that includes the change in concentrations of electrons once the threshold has been reached can be numerically integrated in the general relativistic case. To do that I have used \textit{scipy.integrate} in python and. If $\rho>\bar{\rho}$ then the concentration of electrons at $\beta$ equilibrium is calculated and the pressure is evaluated with the given concentration. The results are in Fig[\ref{fig:tovbeta}].

The value for the limit mass obtained is $M = 1.02146 M_\odot$ and the correspondent $R_\text{Max} = 0.8251R_\text{Earth}$. 

The central pressure at instability is $\rho^{Max}_c = \SI{2.3714e7}{g\per cm^3}$: in accord with the threshold numeric value of beta-decay.

The mass value is significantly different from what found before, this is due to the fact that a three component plasma is only a crude approximation.
\begin{figure}
    \centering
    \includegraphics[width = \columnwidth]{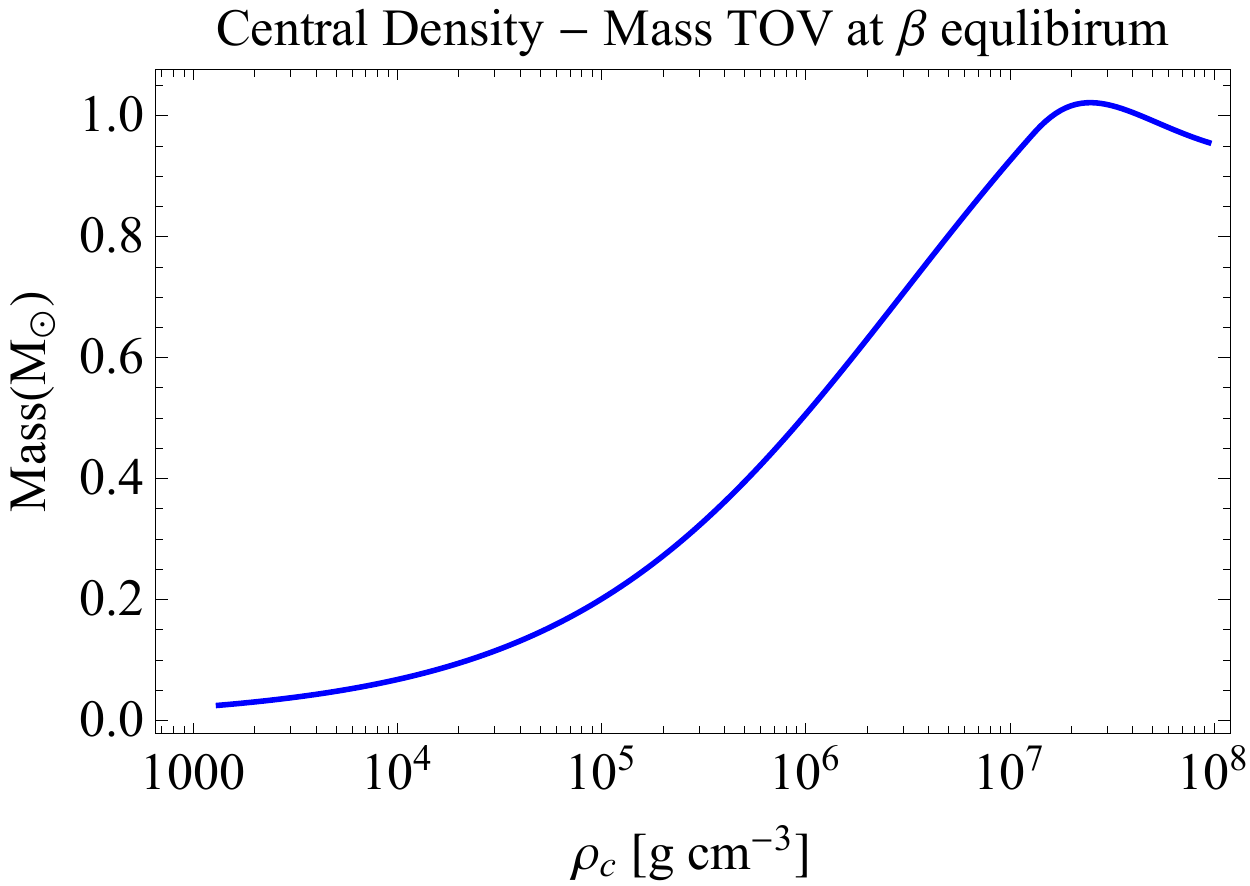}
    \caption{Mass-central density TOV at $beta$-equilibrium, $p,n,e$ mixture}
    \label{fig:centraldenbeta}
\end{figure}

\subsection{A more refined EoS for helium WD}
\label{sec:Helium}
As appointed before a more realistic model should consider the equilibrium between the different compounds of the plasma.

$\beta$-decays in Helium white dwarf involve the main reaction:
\begin{align}
\label{eq:completerea}
    \ce{^4_2He} + e^- \rightarrow \ce{^4_1H} + \nu  \xrightarrow{\text{$10^{-22}$s}}\ce{^3_1 H} + \nu + n 
\end{align}
If we neglect the fact neutrons and \ce{^3_1H} are fermions themselves we can calculate a na\"ive threshold for the Kinetic energy of the electron\cite{shapiro}:
\begin{align*}
    K_F \simeq\SI{20.596}{MeV\per c}
\end{align*}
But for the reaction:
\begin{align}
\label{eq:tritirumdecay}
    \ce{^3_1 H}  + e^- \rightarrow 3n 
\end{align}
I have calculated a threshold of:
\begin{align}
    K_F \simeq \SI{8.752}{MeV\per c}
\end{align}
therefore it is reasonable to consider:
\begin{align}
\label{eq:reactionapprox}
    \ce{^4_2He} + 2e^- \rightarrow 4n + 2\nu
\end{align}
only for purpose of calculating the equilibrium between phases. Even though a three body reaction is particularly rare, we can consider reaction \ref{eq:tritirumdecay} as more rapid than reaction \ref{eq:completerea}, so that \ref{eq:reactionapprox} is a good approximation. 

Because of that, we can consider stellar matter composed by:
\begin{align}
        &\ce{^4_2He} &&e^-&&n
\end{align}

We can therefore study the equilibrium between the 3 phases as in section \ref{sec:beta}. 

As \ce{^4He} is a boson its energy coincide its rest mass in the static approximation. We can write the Baryonic number conservation constraint and electric neutrality as: 
\begin{align}
\label{eq:constrains}
\begin{split}
    &n_{He} + \frac{1}{4}n_n = n \\
    &2n_{He} = n_e  
\end{split}
\end{align}
We want to minimise the total energy, using the Lagrange multipliers method:
\begin{multline}
    F := \epsilon(n_e) + \epsilon(n_n) +n_{He} m_{He}c^2+\\ + \alpha (n_{He} + \frac{1}{4} n_n -n) + \beta (2n_{He} -n_e) 
\end{multline}
We need to impose the the conditions $\frac{\partial F}{\partial n_i}$, derivative have been calculated in the same way of paragraph \ref{sec:beta}:
\begin{align}
\begin{cases}
    \sqrt{(p_ec)^2 + (m_ec^2)^2} = \beta\\
    \sqrt{(p_nc)^2 + (m_nc^2)^2} + \frac{1}{4}\alpha =0 \\
    m_{He} c^2 + \alpha + 2\beta = 0
\end{cases}
\end{align}
So that:
\begin{multline}
    m_{He} c - 4 \sqrt{p_n^2 + (m_n c)^2} +\\ +2\sqrt{p_e^2 + (m_e c)^2} = 0
\end{multline}Together with constraints eq.\ref{eq:constrains} can be solved for $p_e$ for every given Helium density $n$. 

\begin{figure}
    \centering
    \includegraphics[width = \columnwidth]{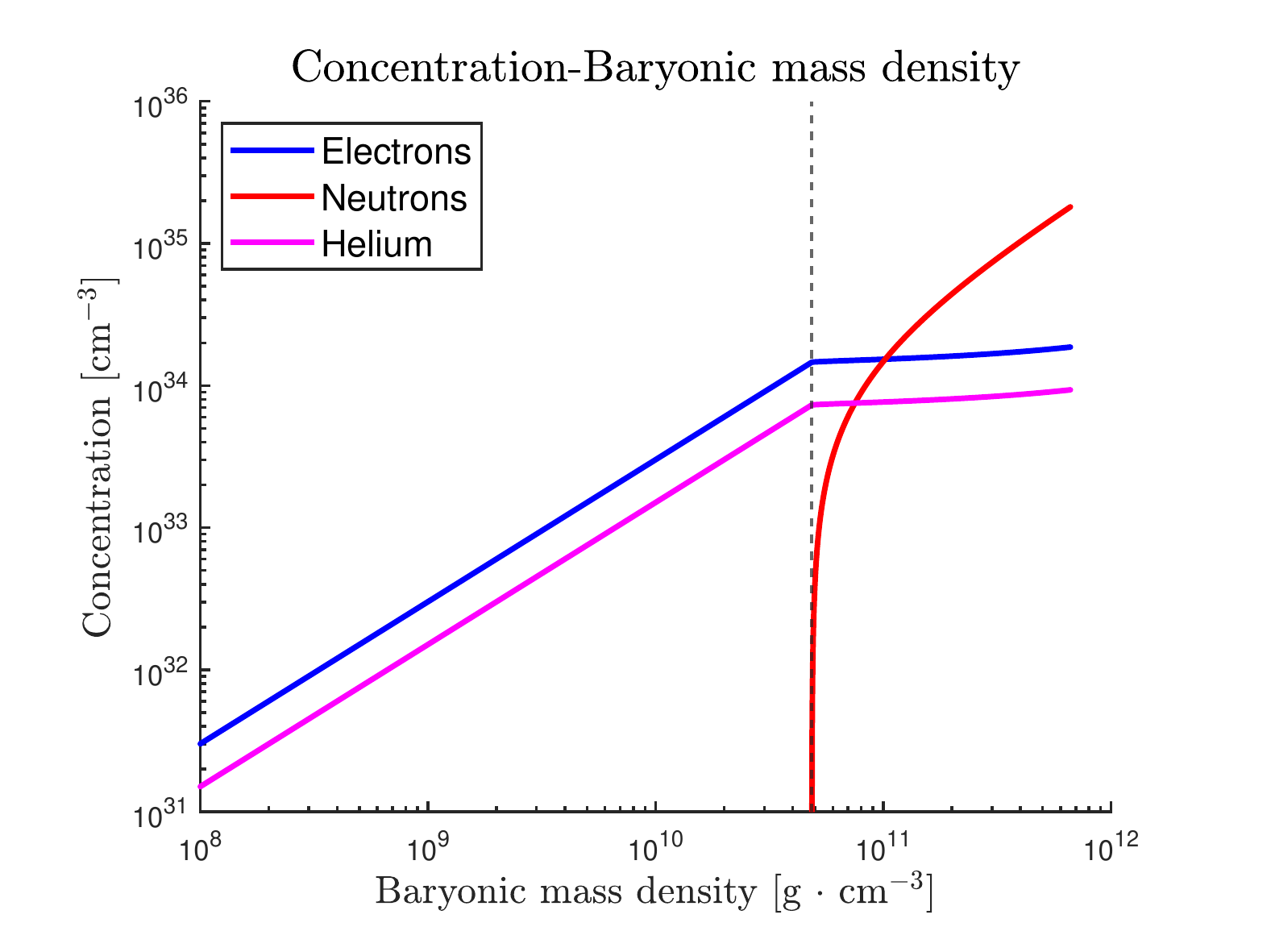}
    \caption{Concentration of $\ce{^4_2He},e^-$ and $n$ at $\beta$-equilibrium}
    \label{fig:conhelium}
\end{figure}

\begin{figure}
    \centering
    \includegraphics[width = \columnwidth]{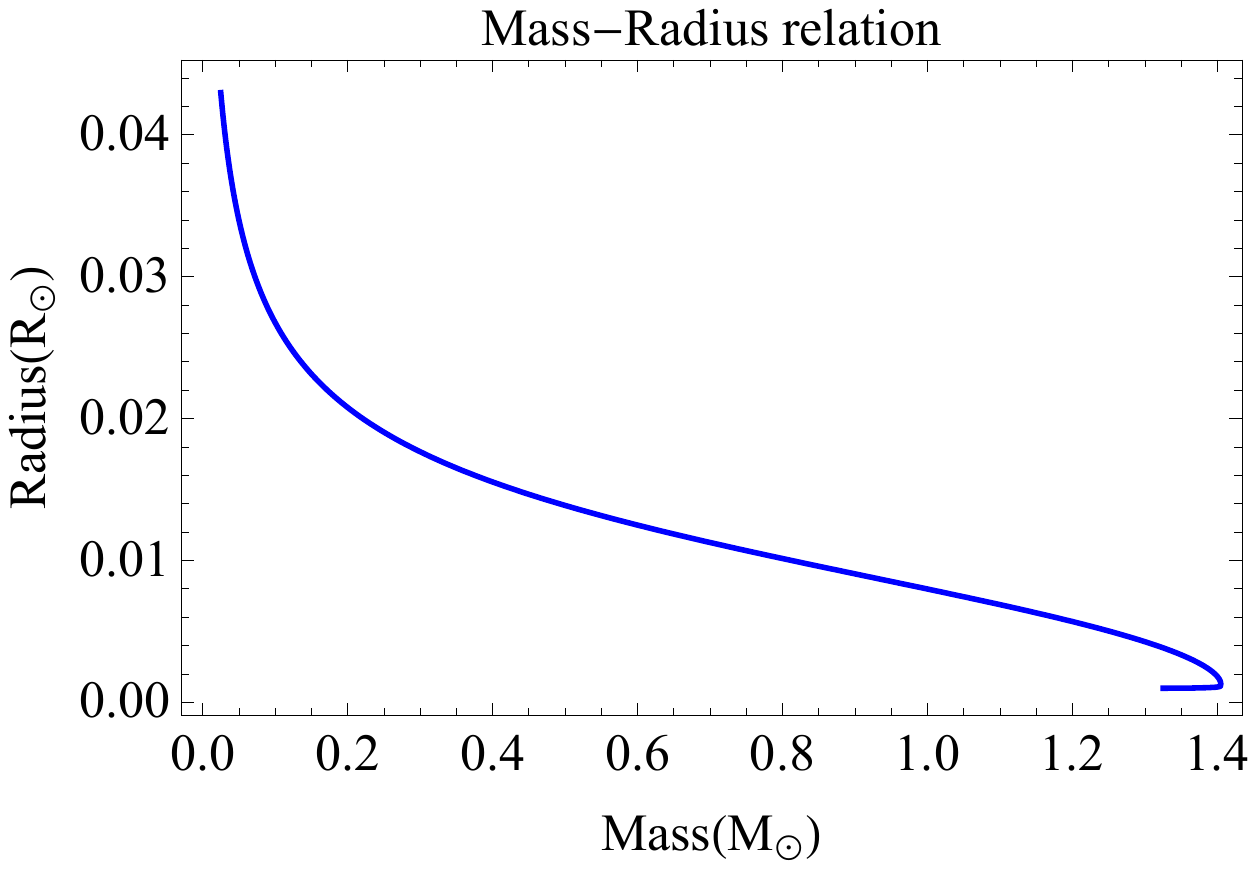}
    \caption{Mass-Radius relation at $\beta$- equilibrium with Helium equation of state}
    \label{fig:tovnewbeta}
\end{figure}

I have computed numerically the EoS. In Fig[\ref{fig:conhelium}] the results obtained. The numeric threshold founded is \begin{align}
\tilde{\rho_c} = \SI{4.8452e+10}{g\per cm^3}\end{align}
As expected $\Tilde{\rho_c}\gg \SI{2.37e+7}{g\per cm^3}$ founded in \ref{sec:beta} as the threshold central density in the case of a mixture of protons neutrons and electrons. 
Thus the EoS can be implemented for integrating the stellar structure equations. In Fig[\ref{fig:tovnewbeta}] the integrated stellar structure with the determined EoS.
With this new EoS the instability is reached first because of the TOV equations rather than because of $\beta$-decay. This is considerable improvement: with the 3 component plasma model of \ref{sec:beta} we had that the instability were reached for $M\ll M_\text{Ch}$.

It would be interesting to repeat the analysis for different chemical compositions. 

\section{Coulomb Interaction and other corrections}
\label{sec:Coulombinteractions}
Coulombian interactions between charged particles have been neglected, in this section the aspect will be deepened. The Salpeter \cite{salpeter} approach will be followed in deriving the results needed. 

\subsection{Classical treatment}
In this first approach quantum effects are neglected and the problem is treated classically. We can assume a perfect lattice of ions and the shape of the Wigner-Seitz cells can be assumed spherical. In a first-order discussion, we neglect also the interactions between the actual cells.

The Wigner-Seitz spherical cell is occupied on average by one single electron. The radius of the cell must be proportional to the Bohr radius $a_0 = \frac{\hbar}{m_e c \alpha}$, we can write it as:
\begin{align}
    R = Z^{1/3}\,\eta\, a_0
\end{align}
The constant of proportionality can be related with the density:
\begin{align}
    n = \frac{\rho}{\mu m_u} = \bigg(\frac{4\pi}{3}(\eta a_0)^3\bigg)^{-1}
\end{align}

It trivial to calculate the energy of a sphere with a charge Ze in the center and a negative uniform distribution -Ze, the potential energy per electron is:
\begin{align}
\label{eq:electrostatic}
    &E_1 = \frac{U}{Z} = -\frac{9}{5}\frac{Z^{1/3}}{\eta} 1\text{Ry} =\\ 
    \label{eq:electrostatic2}
    &=-\frac{9}{10}\frac{Z^{1/3}e^2}{a_0\eta}
\end{align}
The calculation can be found in the Appendix \ref{ap:electr}. 

Given the total energy $E$, the pressure P per electron -at constant chemical potential and constant temperature-  is simply
\begin{align}
    P = -\frac{dE}{dV}= -\frac{d\eta}{dn^{-1}}\frac{dE}{d\eta} 
\end{align}

\paragraph{bcc crystals Madelung constant}
If we were interested in a more precise calculation of the lattice coulombian energy, then we need to use the Madelung constant for a square centered ( bbc ) lattice. In that case instead of $\frac{9}{10}$ in Eq: \ref{eq:electrostatic2}, we need to use $\xi_M=0.895929255682$\cite{Madelung}.
This result is negligible as $\frac{\xi_M-0.9}{0.9} \simeq 0.45\%$. 
\subsection{Perturbative Thomas-Fermi model}
Since effects of coulomb interactions are only small perturbations to the leading Fermi Energy term, we can attempt a perturbative expansion:
\begin{align}
    E = E_0 + E_1 + E_2 + \cdots
\end{align}
$E_0$ is the Fermi energy, $E_1$ is \ref{eq:electrostatic} $E_3$ is often called Thomas-Fermi energy. 

To find $E_3$ we expand the charge distribution n(r) as:
\begin{align}
    n(r) = n_0(1+\lambda(r))
\end{align}
where $n_0$  is the uniform distribution of negative charge . 

If V(r) is the electric potential given by the ion \&  $n_0$. 
We can then determine $n(r)$ imposing that:
\begin{align}
    E_F = mc^2(\sqrt{1+x^2(r)} -1)= c + eV(r)
\end{align}
Where $x = \frac{p_F}{mc}$
This gives an expression for $n(r)$ in terms on c, imposing that $\int_{\mathcal{V}} n(r) dV= n_0$ then c can be calculated. We can now integrate the energy of  $\epsilon(r)$ over the volume and obtain $E_3$. (calculation have been carried out in Appendix \ref{app:TF}):
\begin{align}
\label{eq:TF}
    E_{TF} = -\frac{324}{175}(\frac{4}{9\pi})^{2/3} \sqrt{1+x^2} Z^{4/3} \text{Ry}
\end{align}
\begin{figure}
    \centering
    \includegraphics[width = \columnwidth]{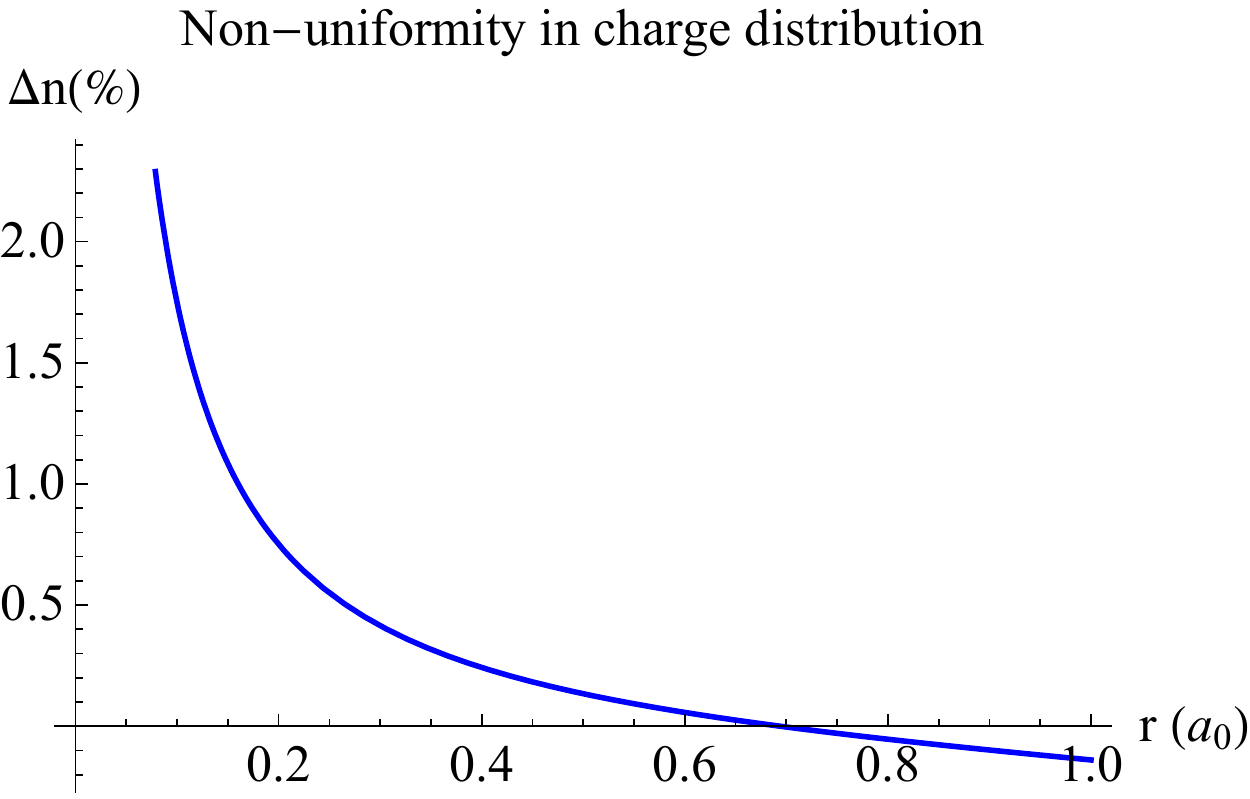}
    \caption{Deviation from uniformity in charge distribution $\Delta n = \frac{n - n_0}{n_0}$ vs radius  (in Bohr Radius $a_0$) for $x=1$}
    \label{fig:nonuni}
\end{figure}
In Fig[\ref{fig:nonuni}] is plotted the deviation from a uniform distribution of charge $\frac{\Delta n(r)}{n_0} = \lambda(r)= \frac{n - n_0}{n_0}$ versus the distance from the central ion (in bohr radius $a_0$) for $x = \frac{P_F}{m_ec} = 1$. It is clear, as expected, that the electrons tend to be concentrated near the ion.
\subsection{Exchange Energy \& correlation energy}
Another effect that need to be considered is the transverse electromagnetic interaction (spin-spin) between electrons. 
The results in cited by Salpeter \cite{salpeter}. The discussion merit more than a single paragraph to be discussed. 

For a complete discussion consult Jancovici\cite{Jancovici}, I cite only the result:
\begin{align}
    E_{ex} = -\frac{3}{4\pi}\alpha m c^2 x \phi(x)
\end{align}
\begin{multline*}
    \phi(x) = \frac{1}{4x^4}\bigg[ \frac{9}{4} + 3 \big(\beta^2 -\frac{1}{\beta^2}\big) \ln\beta \\-6 \ln^2\beta - \big( \beta^2 +\frac{1}{\beta^2}  \big)   -\frac{1}{8}\big(\beta^4 + \frac{1}{\beta^2}\big)   \bigg]
\end{multline*}
\begin{align*}
    \beta = x + \sqrt{1+x^2}
\end{align*}

The next term in the perturbative series is the "Correlation energy": as electrons are fermions exchange of momentum in collisions needs to be considered. This contribution to the energy can be calculated in the approximation of a continuous background of positive charge in which the degenerate electron sea is immersed. 
The problem has been treated by Gell-Mann and Bruecker \cite{gellman}. :
\begin{align}
    E_{corr} = (0.062 \ln \eta - 0.096)ry
\end{align}

\section{Stellar structure with corrections}
\begin{figure}
    \centering
    \includegraphics[width = \columnwidth]{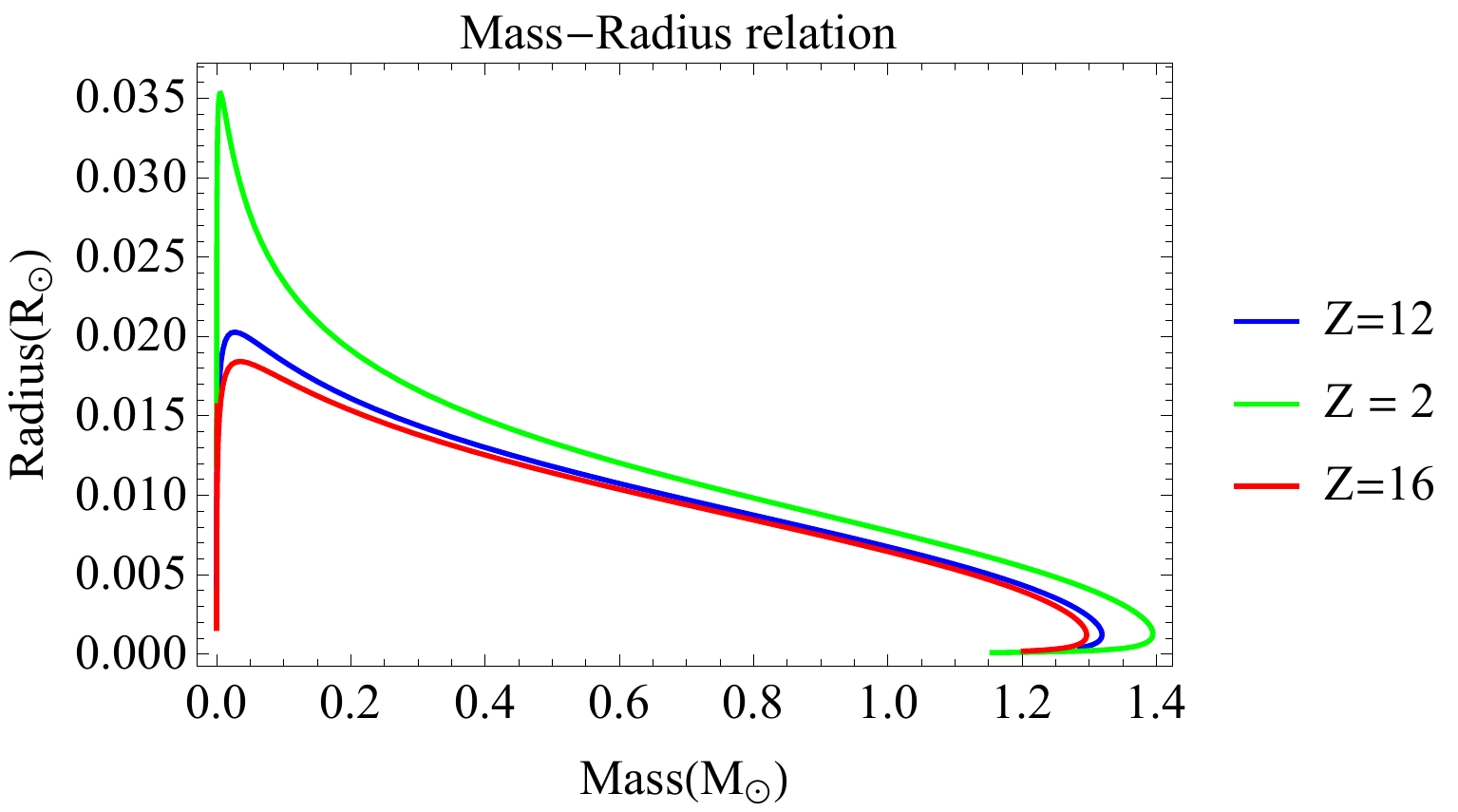}
    \caption{Mass-Radius relation including corrections}
    \label{fig:masssalpeter}
\end{figure}

\begin{figure}
    \centering
    \includegraphics[width = \columnwidth]{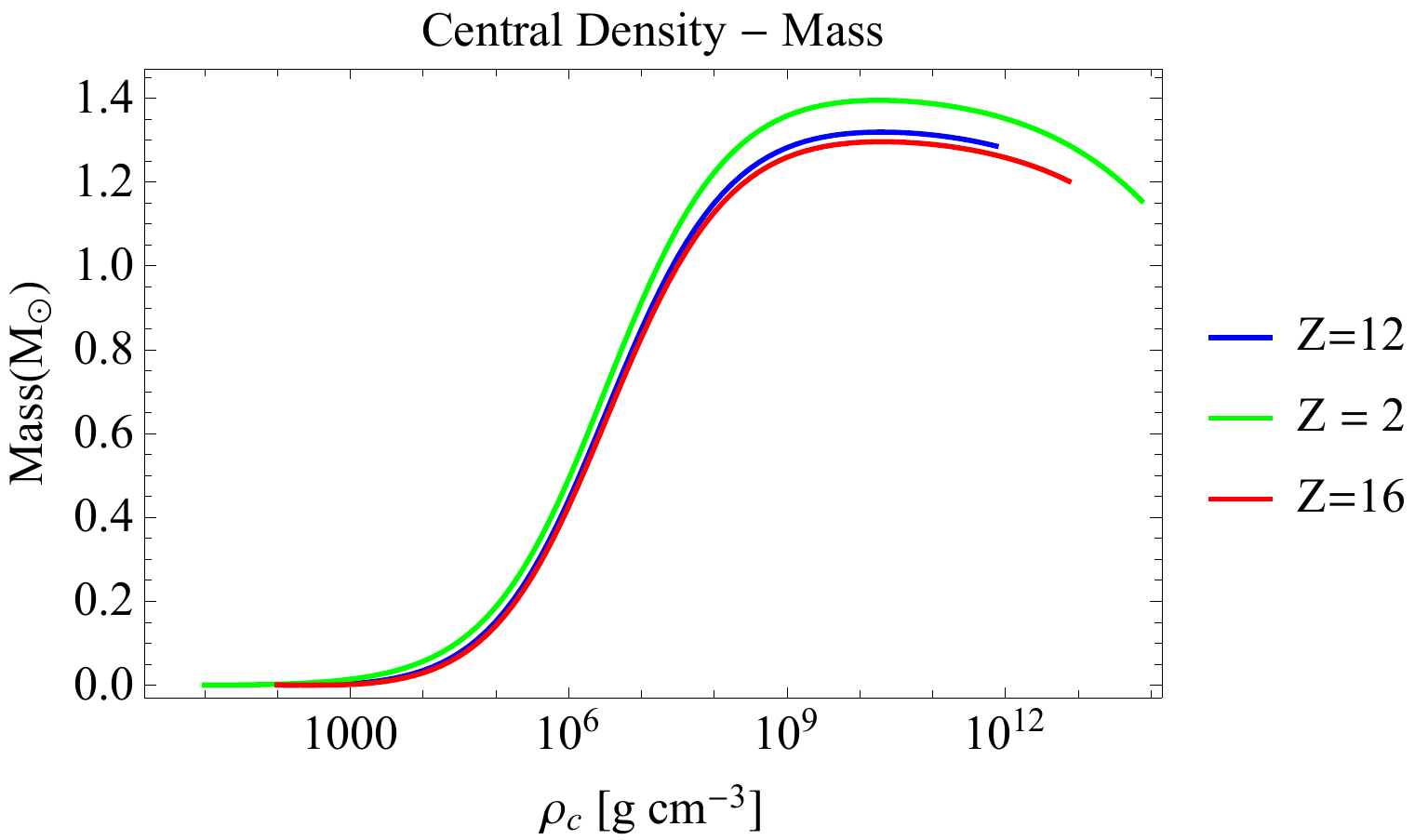}
    \caption{Mass-Central density relation including corrections}
    \label{fig:centralsalpeter}
\end{figure}

\begin{table*}[]
    \centering
    \begin{tabular}{c c c c c}
        \toprule
         Z & $M_{\text{Max}} [M_{\odot}]$ & $R({M_\text{Max}})[R_{\odot}]$ & $\rho(M_{\text{Max}})[\SI{e10}{g \per cm^3}]$&$ R_{\text{Max}}[R_{\odot}]$\\
         \midrule
         2&1.39485&0.001293&1.7783&0.035292 \\
         12&1.31900& 0.001262&1.7783&0.020272\\
         16&1.29607& 0.001254&1.7783&0.018424 \\
         \bottomrule
    \end{tabular}
    \caption{M,R and $\rho_c$ at maximum Mass, and maximum radius of the structure}
    \label{tab:Parameters}
\end{table*}
We have solved numerically the star structure equations in the general relativistic case considering Coulombian interactions and other corrections in section \ref{sec:Coulombinteractions}. $\beta$-decay equilibrium is instated neglected in this calculation.
In Fig[\ref{fig:masssalpeter}] and [\ref{fig:centralsalpeter}] the results obtained. Including Coulomb interaction avoids  divergence at low central density, furthermore it introduces differences due to chemical composition, a more realistic situation. In Tab[\ref{tab:Parameters}] are summarised the relevant feature of the solutions

\begin{figure}
    \centering
    \includegraphics[width = \columnwidth]{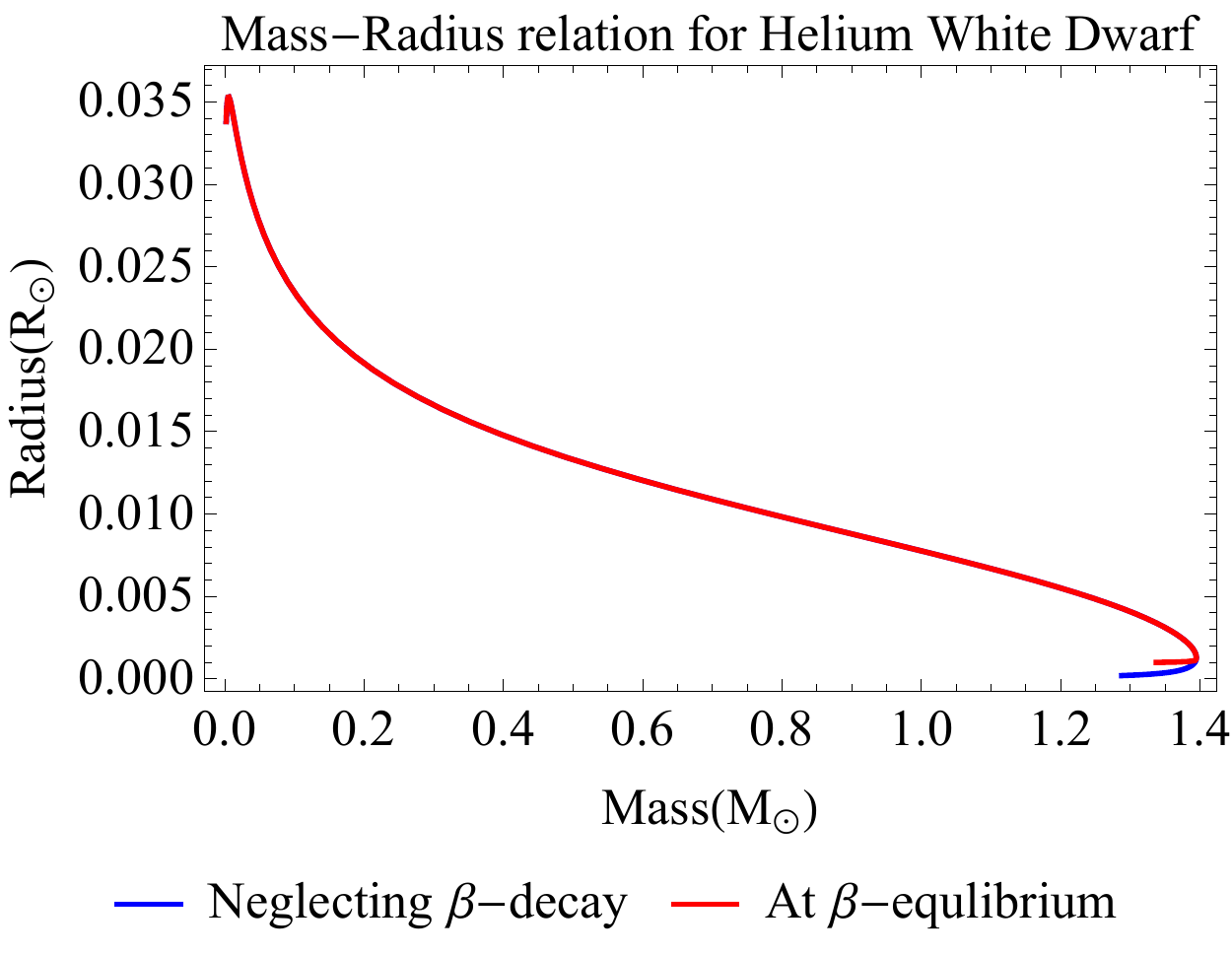}
    \caption{Helium White Dwarf, with and without $\beta$-decay}
    \label{fig:wdhelium}
\end{figure}

\paragraph{Helium WD}
In Fig[\ref{fig:wdhelium}] also $\beta$-decay are considered in the EoS. The model developed in \ref{sec:Helium} is a more realistic one, the $\beta$-decay instability is reached for a more realistic central density.

\section{Conclusion}
The method developed for Helium White dwarf at $\beta$ equilibrium had positive results, in a future work it will be applied also for \ce{^{16}O} and \ce{^{12}C}; effects of finite temperature will be also treated in that work. 

\newpage
\section{Appendix}
\subsection{Evaluation of $\mathcal{E}$ and $P$}
In this section the integrals in Eq. \ref{int:E} and \ref{int:P} are evaluated.
\paragraph{Integral of $P$}
\label{app:intP}
We need to compute:
\begin{align}
    P = \frac{8\pi c}{3h^3}\int_0^{p_F} \frac{p^4}{\sqrt{(mc^2)^2 + (pc)^2}}dp
    \end{align}
Changing to a-dimensional variable first: $\xi = \frac{p}{mc}$ $x = \frac{p_F}{mc}$
\begin{align}
    &P = \frac{8\pi m^4 c^5}{3 h^3}\int_0^x d\xi \frac{\xi^4}{\sqrt{1+\xi^2}} = \\
    & = \frac{8\pi m^4 c^5}{3 h^3}\int_0^{\sinh^{-1}x} \cancel{\cosh{y}} \frac{\sinh^4 y}{\cancel{\cosh{y}}}dy=\nonumber\\
    \label{eq:35}
     &= \frac{8\pi m^4 c^5}{3 \hbar^3}\int_0^{\sinh^{-1}x}\sinh^4ydy=
\end{align}
The integral \ref{eq:35} is straightforward as it can be carried on using the trigonometrical formulas for hyperbolic functions.
\begin{multline}
\label{eq:sinh4}
    \int \sinh^4ydy= \frac{1}{4}\sinh^3{y} \cosh{y} +\\ -\frac{3}{8}\left(\frac{1}{2}\sinh(2y)-x\right)
\end{multline}
Giving \ref{eq:sinh4}, \ref{eq:35} is:
\begin{multline}
=\frac{ m^4 c^5}{24\pi^2 h^3}\bigg[\frac{1}{4}x^3\sqrt{1+x^2}+\\-\frac{3}{8}\left(x\sqrt{1+x^2}-\sinh^{-1}x\right)\bigg]
\end{multline}
Which is \ref{int:P}.
\paragraph{Integral of $\mathcal{E}$}
Computing \ref{int:E} is easy once \ref{int:P} has been computed, in fact 
\label{sec:integrals}
\begin{align}
    &\mathcal{E} = \frac{1}{8\pi^2\hbar^3}\int_0^{p_F} p^2\sqrt{(mc^2)^2 + (pc)^2}dp =\nonumber\\ &= \frac{m^4 c^5}{8\pi^2\hbar^3}\int_0^x \xi^2\sqrt{1+\xi^2}d\xi =\nonumber \\
    \label{eq:integraldone}
    &= \frac{m^4 c^5}{24\pi^2\hbar^3} \bigg[x^3\sqrt{1+x^2} - \int_0^x \frac{\xi^4}{\sqrt{1+\xi^2}} d\xi\bigg]
\end{align}
The integral in \ref{eq:integraldone} was evaluated in \ref{app:intP}, hence we obtain \ref{int:E}.
\subsection{Electrostatic energy}
\label{ap:electr}
The total energy is given by the sum of the of interaction between the ion and the electronic cloud and the energy of the spherical distributing itself 
\begin{align}
&E_{TOT} = E_{Sph} + E_{int} =\\&= \frac{3}{5}\frac{(Ze)^2}{R} - \int_0^R \bigg(\frac{Ze}{r}\bigg) \frac{Ze}{\frac{4}{3}\pi R^3}4\pi r^2 dr = \\
&=(\frac{3}{5} - \frac{3}{2}) \frac{(Ze)^2}{R} = -\frac{9}{5}\frac{(Ze)^2}{R} 
\end{align}
Eq. \ref{eq:electrostatic} is obtained with $R\rightarrow Z^{1/3}\eta a_0$ and dividing by Z.
\subsection{Thomas-Fermi Energy}
\label{app:TF}
\begin{align}
    n(r) = n_0 (1+ \lambda(r))
\end{align}
The Fermi Energy is:
\begin{align}
    &E_F = m_ec^2\sqrt{1+x^2(r)} -1 \\
    &x(r) = \frac{(3\pi^2\hbar^3)^{1/3}}{m_e c} n^{1/3}(r)
\end{align}
We need to impose that at a first order of expansion in $\lambda$:
\begin{align}
    E_F = c + V(r) 
\end{align}
If,
\begin{align*}
    H = \left(\frac{3n_0\pi^2\hbar^3}{(mc^2)^3}\right)^{2/3}
\end{align*}
Then $E_F$ can be expanded in a neighbourhood of $\tilde{E_F}=E_F\left(n(r) = n_0\right)$ as:
\begin{align}
    &E_F = mc^2\bigg[\left(1+ H\left(1+\lambda\right)^{2/3}\right)^{1/2}-1\bigg] =\nonumber\\
    \label{eq:fermienergy}
    &= \tilde{E_F} + \frac{H mc^2}{3\sqrt{1+H}} \lambda + o(\lambda^2) 
\end{align}
We now need to evaluate the Potential $V(r)$ due to a uniform distribution of negative charge $-Ze$ in a uniform sphere of radius R  $n_0 = \frac{-3Ze}{4\pi R^2}$, this is trivial:
\begin{align}
    V_-(r)=\begin{cases}
    \frac{4}{3}\frac{\pi}{2}n_0(3R^2 -r^2) &r\leq R\\
    \frac{4}{3}\pi n_0\frac{R^3}{r} = -\frac{Ze}{r} &r>R
    \end{cases}
\end{align}
So that the total potential is:
\begin{align}
V(r) = \begin{cases}
    \frac{2\pi n_0}{3}(3R^2 -r^2) +\frac{Ze}{r} &r<R\\
    0 & r>R
\end{cases}
\end{align}
We can therefore find the value of $\lambda(r)$ by imposing that eq- \ref{eq:fermienergy} is equal to the potential:
\begin{align}
    \lambda  = \left(c-\tilde{E_F} + eV(r)\right)\frac{3\sqrt{1+H}}{Hmc^2}
\end{align}
Where c is a constant.
To find c, we impose that:
\begin{align}
\int_0^R \lambda(r) r^2 dr= 0
\end{align}

In that way an analytic expression for $\lambda$ is obtained, in Fig[\ref{fig:nonuni}] is plotted $\lambda$ in percentage. 

In the end, to find the corrections in energy due to the non uniformity, we integrate:
\begin{align}
   E_{TF} = \int_0^R 4\pi V(r) n_0 \lambda(r) r^2 dr 
\end{align}
Doing that the eq \ref{eq:TF} is obtained expressing R in function of $x$.

\medskip
\bibliographystyle{alpha}
\bibliography{biblio}

\end{document}